\documentclass{aa}  

\usepackage{graphicx}
\usepackage{txfonts}
\usepackage{amsmath}

\usepackage{kantlipsum} 
\usepackage{microtype}
\usepackage{soul}

\usepackage[colorlinks=true]{hyperref}
\hypersetup{citecolor= blue, linkcolor=red}

\usepackage{color}

\newcommand{\enhance}{{\tt Enhance}}

\begin{document} 

   \title{Enhancing SDO/HMI images using deep learning}
   \author{C. J. D\'{\i}az Baso
          \inst{1,2}
          \and
          A. Asensio Ramos\inst{1,2}
          }

   \institute{Instituto de Astrof\'isica de Canarias, C/V\'{\i}a L\'actea s/n, E-38205 La Laguna, Tenerife, Spain
   \and
   Departamento de Astrof\'{\i}­sica, Universidad de La Laguna, E-38206 La Laguna, Tenerife, Spain
             }

   \date{}
   
   \titlerunning{Enhancing HMI images}

 
  \abstract
   {The Helioseismic and Magnetic Imager (HMI) provides continuum images and
   magnetograms with a cadence better than one every minute. It has been continuously observing the Sun 
   24 hours a day for the past 7 years. The trade-off between full disk observations and spatial resolution makes that HMI is not enough to analyze the smallest-scale
   events in the solar atmosphere.}
   {Our aim is developing a new method to enhance HMI data, simultaneously deconvolving
   and superresolving images and magnetograms. The resulting images will mimick
   observations with a diffraction-limited telescope twice the diameter of HMI.}
   {The method, that we term \enhance, is based on two deep fully convolutional 
   neural networks that input patches of HMI observations and output
   deconvolved and superresolved data. The neural networks are trained on
   synthetic data obtained from simulations of the emergence of solar
   active regions.}
   {We have obtained deconvolved and supperresolved HMI images. To solve this ill-defined problem 
   with infinite solutions we have used a neural network approach to add prior information 
   from the simulations.
   We test \enhance\ against Hinode data that has been degraded to a 28 cm diameter
   telescope showing very good consistency. The code is open sourced for the community.}
   {}


   \maketitle

\microtypesetup{activate=true}


\section{Introduction}
Astronomical observations from Earth are always limited
by the presence of the atmosphere, which strongly disturbs
the images. An obvious (but expensive) solution to this problem is to place the
telescopes in space, which produces observations
without any (or limited) atmospheric aberrations. Although the observations
obtained from space are not affected by atmospheric seeing, the optical
properties of the instrument still limits the observations.

In the case of near-diffraction limited observations, the
point spread function (PSF) establishes the maximum allowed
spatial resolution. The PSF typically contains two different
contributions. The central core is usually
dominated by the Airy diffraction pattern, a consequence
of the finite and circular aperture of the telescope (plus other perturbations
on the pupil of the telescope like the spiders used to keep
the secondary mirror in place). The tails of the PSF are
usually dominated by uncontrolled sources of dispersed light
inside the instrument, the so-called stray light. It is known
that the central core limits the spatial resolution of the
observations (the smallest feature that one can see in the image), while
the tails reduce the contrast of the image \citep{Danilovic2010}. Moreover, 
it is important to note
that knowing the PSF of any instrument is a very complicated
task \citep{Yeo2014,Couvidat2016}.

If the PSF is known with some precision, it is possible to apply
deconvolution techniques to partially remove the perturbing
effect of the telescope. The deconvolution is usually
carried out with the Richardson-Lucy algorithm \citep[RL;][]{richardson72},
an iterative procedure that returns a maximum-likelihood solution to the problem.
Single image deconvolution is usually
a very ill-defined problem, in which a potentially infinite number of solutions can be 
compatible with the observations. Consequently, some kind of regularization
has to be imposed. Typically, an early-stopping strategy in the iterative process of the RL algorithm
leads to a decent output, damping the high spatial frequencies
that appear in any deconvolution process. However, a maximum a-posteriori approach in which
some prior information about the image is introduced gives much
better results.

Fortunately, spectroscopic and spectropolarimetric observations provide
multi-image observations of a field-of-view (FOV) and the deconvolution
process is much better defined. This deconvolution process has been tried recently with
great success by \cite{vannoort12}, who also introduced a strong
regularization by assuming that the Stokes profiles in 
every pixel have to be explained with the emerging Stokes profiles
from a relatively simple model atmosphere assuming local thermodynamical
equilibrium. Another solution was provided by \cite{ruizcobo_asensioramos13},
who assumed that the matrix built with the Stokes profiles for all
observed pixels has very low rank. In other words, it means that the Stokes 
profiles on the FOV can be linearly expanded with a reduced set of vectors. 
This method was later exploited 
by \cite{Quintero2015} with good results. 
Another different approach
was developed by \cite{Asensio2015} where they used the concept of sparsity (or compressibility),
which means that one can linearly expand the unknown quantities in a basis set with only a 
few of the elements of the basis set being active. Under the assumption of sparsity, they 
exploited the presence of spatial correlation on the maps of physical parameters, 
carrying out successful inversions and deconvolution simultaneously.

A great science case for the application of deconvolution and 
superresolution techniques is the 
Helioseismic and Magnetic Imager \citep[HMI;][]{Scherrer2012} 
onboard the Solar Dynamics Observatory \citep[SDO;][]{sdo2012}.
HMI is a space-borne observatory that deploys 
full-disk images (plus a magnetogram and dopplergram) of the Sun 
every 45 s (or every 720 s for a better signal-to-noise ratio).
The spatial resolution of these images is $\sim 1.1''$,
with a sampling of $\sim 0.5''$/pix. In spite of the enormous
advantage of having such a synoptic spatial telescope without the 
problematic earth's atmosphere, the spatial resolution is not enough 
to track many of the small-scale solar structures of 
interest. The main reason of that is the sacrifice that HMI does to cover 
the full disk of the Sun encapsulating that FOV on a feasible sensor.
We think that, in the process of pushing for the science advance, 
one would desirably prefer images with a better spatial resolution 
and compensated for the telescope PSF. 

Under the assumption of the linear theory of image formation, and writing images 
in lexicographic order (so that they are assumed to be sampled at a given 
resolution), the observed image can be written as:
\begin{equation}
\mathbf{I} = \mathbf{D} [\mathbf{P} * \mathbf{O}] + \mathbf{N}, 
\end{equation}
where $\mathbf{O}$ is the solar image at the entrance of the
telescope, $\mathbf{P}$ is a convolution matrix that simulates the effect
of the PSF on the image, $\mathbf{D}$ is a sub-sampling (non-square) matrix that reduces the
resolution of the input image to the desired
output spatial resolution and $\mathbf{N}$ represents noise (usually with Gaussian or Poisson statistics).
The solution to the single-image deconvolution+superresolution problem \citep[SR;][]{Borman1998} requires
the recovery of $\mathbf{O}$ (a high-resolution image of $2N \times 2N$ pixels) from a single measurement 
$\mathbf{I}$ (a low-resolution image of $N \times N$ pixels). 
This problem is extremely ill-posed, even worse than the usual deconvolution
to correct from the effect of the PSF. A multiplicity (potentially an infinite number)
of solutions exists. This problem is then typically solved by imposing
strong priors on the image \citep[e.g.,][]{Tipping03}.

Despite the difficulty of the problem, we think there is 
great interest in enhancing the HMI images using post-facto
techniques. A super-resolved image could help detect 
or characterize small features in the surface of the Sun, or improve
the estimation of the total magnetic flux limited by the resolution 
in the case of magnetograms.
This motivated us to develop an end-to-end fast method
based on a deep fully convolutional neural network 
that simultaneously deconvolve and superresolve by a factor of 2 
the HMI continuum images and magnetograms. We have preferred to be conservative and only do superresolution by a factor 2 because 
our tests with a larger factor did not produced satisfactory results.
Deep learning single-image deconvolution and superresolution has been recently applied with
great success in natural images \citep{Xu2014,Dong2015,Dong2016,Shi2016,Ledig2016,Hayat2017}. Given 
the variability of all possible natural images, a training-based
approach should give much better results in our case than in the 
case of natural images. 
In the following, we give details about the
architecture and training of the neural network and provide examples
of applications to HMI data.


\begin{figure*}
\centering
\includegraphics[width=0.4\textwidth,trim={0 0 0 0},clip]{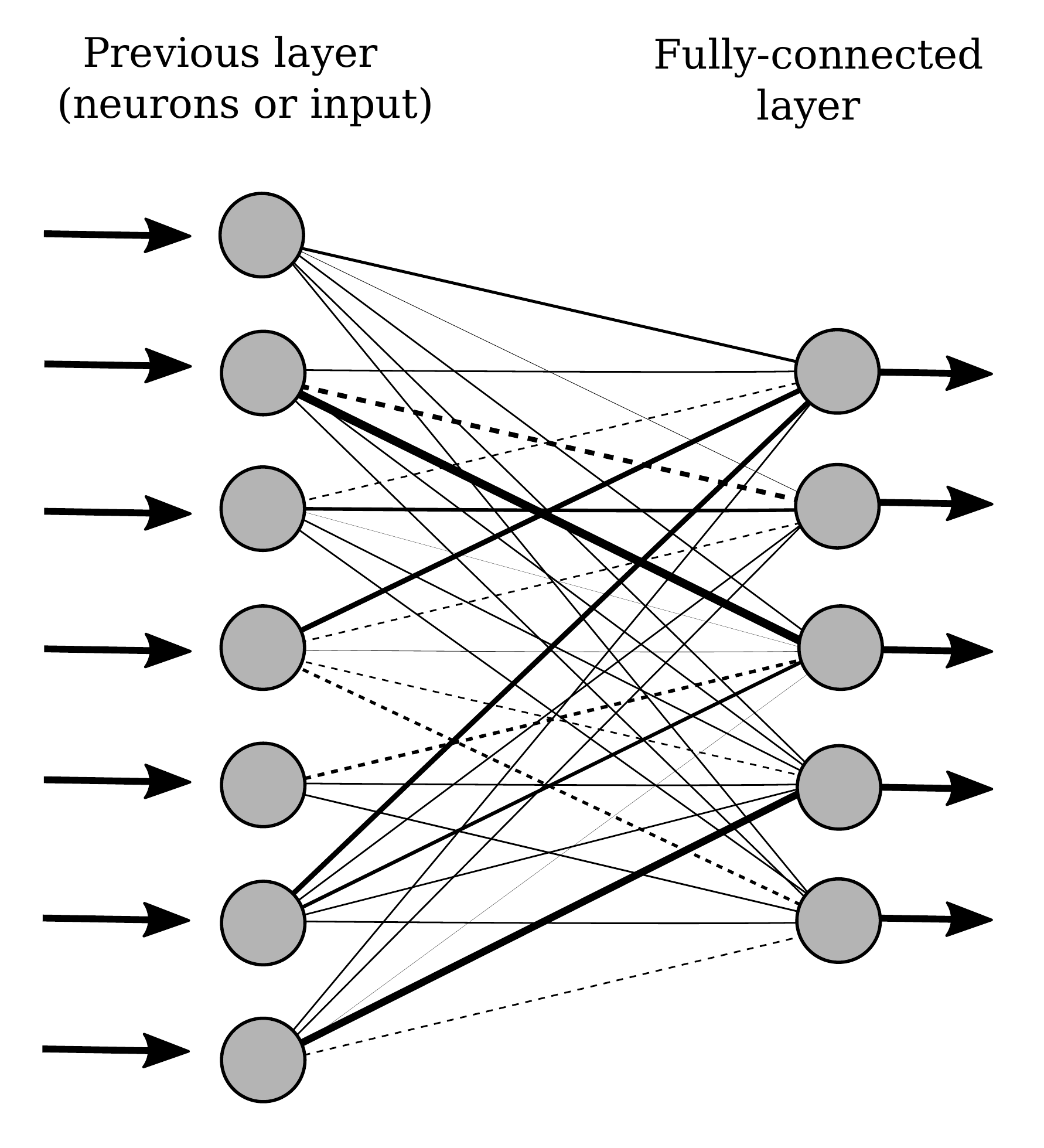}
\hspace{5pt} \vrule width 0.5pt \hspace{5pt}
\includegraphics[width=0.5\textwidth,trim={0 0 0 0},clip]{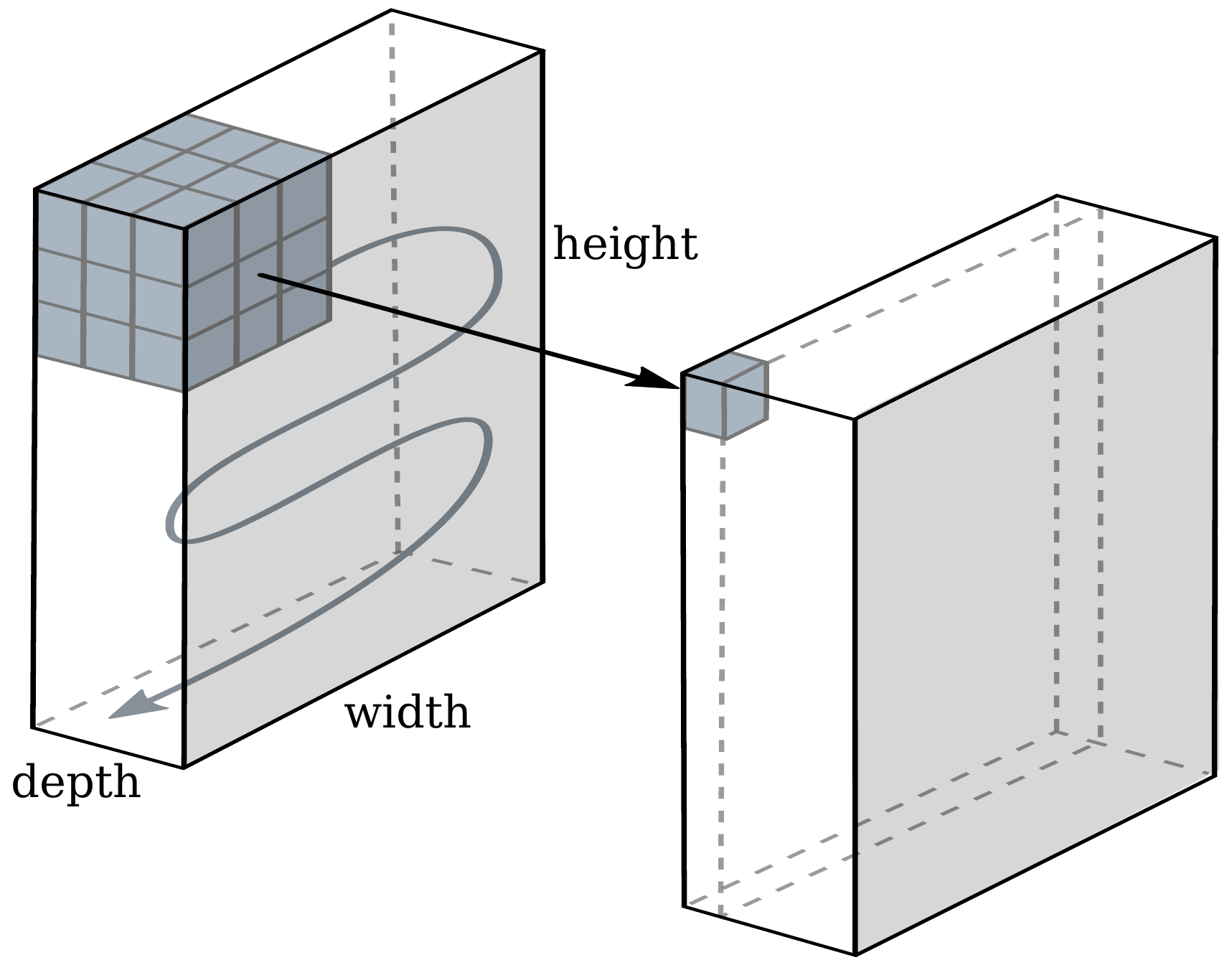}
\caption{Left panel: building block of a fully-connected neural network. Each input of the previous 
layer is connected to each neuron of the output. Each connection is represent by different 
lines where the width is associated to higher weights and the dashed lines to negative weights. 
Right panel: three-dimensional convolution carried out by a convolutional layer. The 3D-kernel traverses
the whole input, producing a single scalar at each position. At the end, a 2D feature map will be 
created for each 3D kernel. When all feature maps are stacked, a feature map tensor will be created.}
\label{fig:networks}
\end{figure*}

\section{Deep Convolutional Neural Network}

\subsection{Deep neural networks}

Artificial neural networks (ANN) are well-known computing systems based on connectionism
that can be considered to be very powerful approximants to arbitrary functions \citep{B96}.
They are constructed by putting together many basic fundamental structures (called neurons)
and connecting them massively. Each neuron $i$ is only able to carry out a very basic operation
on the input vector: it multiplies all the input values $x_j$ by some weights $w_j$, 
adds some bias $b_i$ and finally returns the value of a certain user-defined
nonlinear activation function $f(x)$. In mathematical notation, a neuron computes: 
\begin{equation}
o_i = f(\Sigma_j\,x_j\cdot w_j + b_i).
\end{equation}
The output $o_i$ is then input in another neuron that does a similar work.
 
An ANN can be understood as a pipeline where the information goes from the input to the output, 
where each neuron makes a transformation like the one described above (see left panel of Fig. 
\ref{fig:networks}). Given that neurons are usually grouped in layers, the term deep neural network 
comes from the large number of layers that are used to build the neural network. Some of 
the most successful and recent neural networks contain several millions of neurons organized in
several tens or hundreds of layers \citep{veryDeep2014}. As a consequence, deep neural networks can 
be considered to be a very complex composition of very simple nonlinear functions, which gives 
the capacity to do very complex transformations.

The most used type of neural network from the 1980s to the 2000s
is the fully connected network \citep[FCN; see][for an overview]{Overview2014}, 
in which every input is connected to every neuron of the following layer. Likewise, the output 
transformation becomes the input of the following layer (see left panel of Fig. \ref{fig:networks}).
This kind of architecture succeeded to solve problems that were considered to be not easily solvable as the recognition of handwritten characters \citep{B96}. A selection of applications in Solar Physics include the inversion of Stokes profiles 
\citep[e.g.,][]{socas05,carroll08}, the acceleration of the solution of chemical equilibrium \citep{asensio05}
and the automatic classification of sunspot groups \citep{colak08}.

Neural networks are optimized iteratively by updating the weights and biases so that 
a loss function that measures the ability of the network to predict the output from the input 
is minimized\footnote{This is the case of supervised training. Unsupervised neural networks
are also widespread but are of no concern in this paper.}. This optimization is widely known as 
learning or training process. In this process a training dataset is required.

\subsection{Convolutional neural networks}
In spite of the relative success of neural networks, their application to 
high-dimensional objects like images or videos turned out to be
an obstacle. The fundamental reason was that the number of
weights in a fully connected network increases extremely fast with the complexity of the network (number of neurons) and the computation quickly becomes unfeasible.
As each neuron has to be connected with the whole input, if we add a new neuron we will add the size of the input in number of weights. Then, a larger number of neurons implies a huge number of connections.
This constituted an
apparently unsurmountable handicap that was only solved with the
appearance of convolution neural networks \citep[CNN or ConvNets;][]{LeCun1998}.

The most important ingredient in the CNN is the convolutional layer which is composed of
several convolutional neurons. Each CNN-neuron carries out the convolution of the input with a certain
(typically small) kernel, providing as output what is known as feature map. Similar to a FCN, the output of
convolutional neurons is often passed through a nonlinear activation function.
The fundamental advantage of CNNs is that the same weights are shared across the whole input, 
drastically reducing the number of unknowns. This also makes
CNN shift invariant (features can be detected in an image irrespectively of
where they are located). 


In mathematical notation, for a two-dimensional input $X$
of size $N \times N$ with $C$ channels\footnote{The term channels is inherited from
the those of a color image (e.g., RGB channels). However, the term has a much more general
scope and can be used for arbitrary quantities \citep[see][for an application]{Asensio2017}.} 
(really a cube or tensor of size $N \times N \times C$), each output feature map $O_i$ (with size $N \times N \times 1$) of a convolutional layer is computed as:
\begin{equation}
O_i=K_i * X + b_i,
\end{equation}
where $K_i$ is the $K \times K \times C$ kernel tensor associated with the output feature map $i$, 
$b_i$ is a bias value ($1 \times 1 \times 1$) and the convolution is displayed with the symbol $*$. 
Once the convolution with $M$ different kernels is carried out and stacked together, the output 
$O$ will have size $N \times N \times M$. All convolutions are here indeed intrinsically three dimensional, 
but one could see them as the total of $M \times C$ two dimensional convolutions plus the 
bias (see right panel of Fig. \ref{fig:networks}).

\begin{figure*}
\centering
\includegraphics[width=\textwidth]{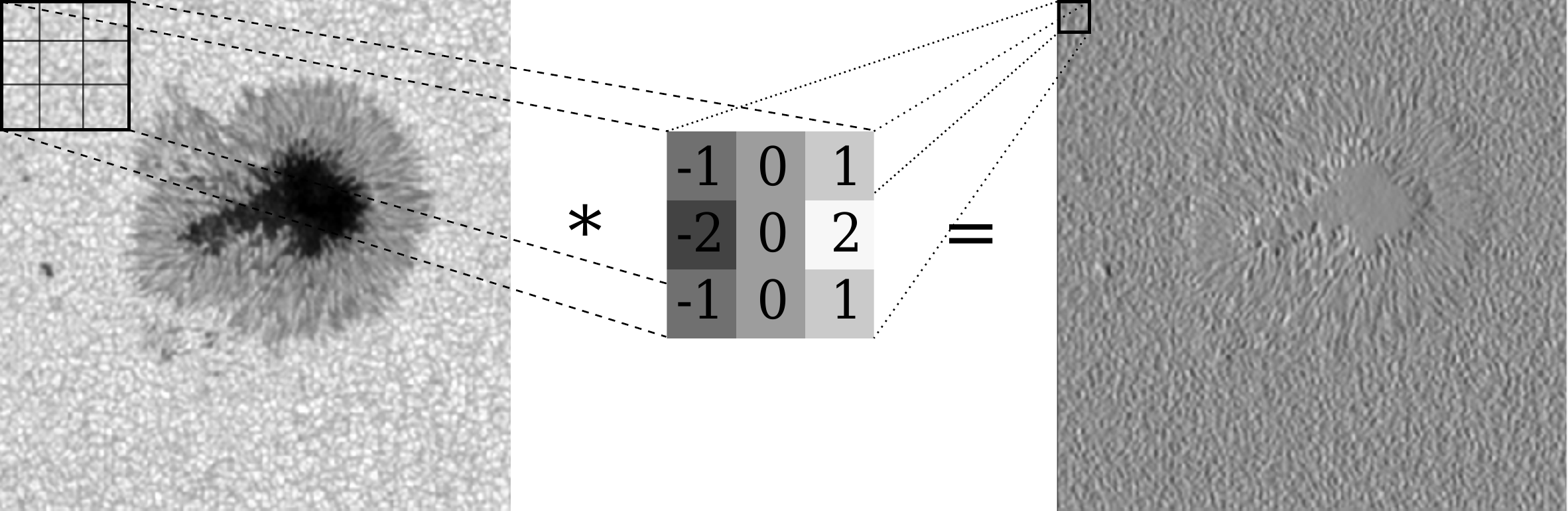}
\caption{An example of a convolution with a filter. In this example a vertical border-locating
kernel is convolved with the input image of the Sun. A resulting feature map of size
$(N-2)\times(N-2)$ is generated from the convolution.}
\label{fig:featuremap}
\end{figure*}

CNNs are typically composed of several layers. This layerwise architecture exploits the 
property that many natural signals are a generated by a hierarchical composition of 
patterns. For instance, faces are composed of eyes, while eyes contain a similar internal structure. 
This way, one can devise specific kernels that extract this information from 
the input. As an example, Fig. \ref{fig:featuremap} shows the effect of a vertical border detection 
kernel on a real solar image. The result at the right of the figure is the feature map.  
CNNs work on the idea that each convolution layer extracts information about certain patterns, 
which is done during the training by iteratively adapting the set of convolutional
kernels to the specific features to locate. This obviously leads to a much more optimal solution as
compared with hand-crafted kernels. Despite the exponentially smaller 
number of free parameters as compared with a fully-connected ANN, CNNs produce much better 
results. It is interesting to note that, 
since a convolutional layer just computes sums and multiplications of the inputs, 
a multi-layer FCN (also known as perceptron) is perfectly capable of reproducing it, but it would require more 
training time (and data) to learn to approximate that mode of operation 
\citep{Peyrard15}.



Although a convolutional layer significantly decreases 
the number of free parameters as compared with a fully-connected layer, it 
introduces some hyperparameters (global characteristics of the network) to be set in 
advance: the number of kernels
to be used (number of feature maps to extract from the input), size of 
each kernel with its corresponding padding (to deal with the borders of the image)
and stride (step to be used during the convolution
operation) and the number of convolutional layers and specific architecture to 
use in the network. As a general rule, the deeper the CNN, the better the result, 
at the expense of a more difficult and computationally intensive training. CNNs
have been used recently in astrophysics for denoising images of galaxies 
\citep{schawinski17}, for cosmic string detection in CMB temperature 
maps \citep{Ciuca17}, or for the estimation of horizontal velocities 
in the solar surface \citep{Asensio2017} .

\subsection{Activation layers}
As said, the output of a convolutional layer is often passed through a non-linear function 
that is termed the activation function. Since the convolution operation is linear, this activation
is the one that introduces the non-linear character of the CNNs. Although hyperbolic
tangent, $f(x)=\tanh(x)$, or sigmoidal, $f(x)=[1+\exp(-x)]^{-1}$, activation units were originally 
used in ANNs, nowadays a panoply of more convenient nonlinearities are used. The main problem 
with any sigmoid-type activation
function is that its gradient vanishes for very large values, difficulting
the training of the network. Probably the most common activation function is 
the rectified linear unit \citep[ReLU;][]{relu10} or slight variations of it. The ReLU
replaces all negative values in the input by zero and keeps the rest untouched. This activation
has the desirable property of producing non-vanishing gradients for positive
arguments, which greatly accelerates the training.

\subsection{General training process}

CNNs are trained by iteratively modifying the weights and biases of the convolutional 
layers (and any other possibly learnable parameter in the activation layer). The aim
is to optimize a user-defined loss function from the output
of the network and the desired output of the training data. The optimization
is routinely solved using simple first-order gradient descent algorithms \citep[GD; see][]{Rumelhart1988}, which modifies the weights along
the negative gradient of the loss function with respect to the model parameters
to carry out the update. The gradient of the loss function with respect to the 
free parameters of the neural network is obtained through the
backpropagation algorithm \citep{LeCun1998b}. Given that neural networks are 
defined as a stack of modules (or layers), the gradient of the loss 
function can be calculated using the chain rule as the product of the 
gradient of each module and, ultimately, of the last layer and the
specific loss function.

In practice,
procedures based on the so-called stochastic gradient descent (SGD) are used, in which
only a few examples (termed batch) from the training set are used
during each iteration to compute a noisy estimation of the gradient and adjust the weights 
accordingly. Although the calculated gradient is a noisy estimation of the one calculated with the whole training set, the training is faster as we have less to compute and
more reliable. If the general loss function $Q$ is the average of each loss $Q_j$  computed on a batch of inputs and it can be written as $Q=\Sigma_j^n Q_j/n$, the weights $w_i$ are updated following the same recipe as the 
gradient descend algorithm but calculating the gradient within a single batch:
\begin{equation}
w_{i+1} = w_i -\eta\nabla Q(w_i) = w_i -\eta\nabla\Sigma_j^n Q_j(w_i)/n \simeq w_i -\eta\nabla Q_j(w_i),
\end{equation}
where $\eta$ is the so-called learning rate.
It can be kept fixed or it can be changed according to our requirements. This parameter has to be tuned to find a compromise between the accuracy of the network and the speed of convergence. If $\eta$ is too large, the steps will be too large and the solution could 
potentially overshoot the minimum. On the contrary, if it is too small it will take so many iterations to reach the minimum. Adaptive methods like
Adam \citep{adam14} have been developed to automatically tune the
learning rate.


Because of the large number of free parameters in a deep CNNs, overfitting can be 
a problem. One would like the network to generalize well and avoid any type of "memorization" of
the training set. To check for that, a part of the training set is not used during the update
of the weights but used after each iteration as validation. Desirably, the loss should
decrease both in the training and validation sets simultaneously. If overfitting occurs, the
loss in the validation set will increase.

Moreover, several techniques have been described in the literature to accelerate
the training of CNNs and also to improve generalization. Batch normalization \citep{batch_normalization15}
is a very convenient and easy-to-use technique that consistently produces large accelerations in the
training. It works by normalizing every batch to have 
zero mean and unit variance. Mathematically, the input is normalized so that:
\begin{align}
y_i &= \gamma \hat{x_i} + \beta \nonumber \\
\hat{x_i} &= \frac{x_i - \mu}{\sqrt{\sigma^2 + \epsilon}},
\end{align}
where $\mu$ and $\sigma$ are the mean and standard deviation of the inputs on the batch and
$\epsilon=10^{-3}$ is a small number to avoid underflow. The parameters $\gamma$ and $\beta$
are learnable parameters that are modified during the training.

\begin{figure*}
\centering
\includegraphics[width=0.9\textwidth,trim={0 0 0 0},clip]{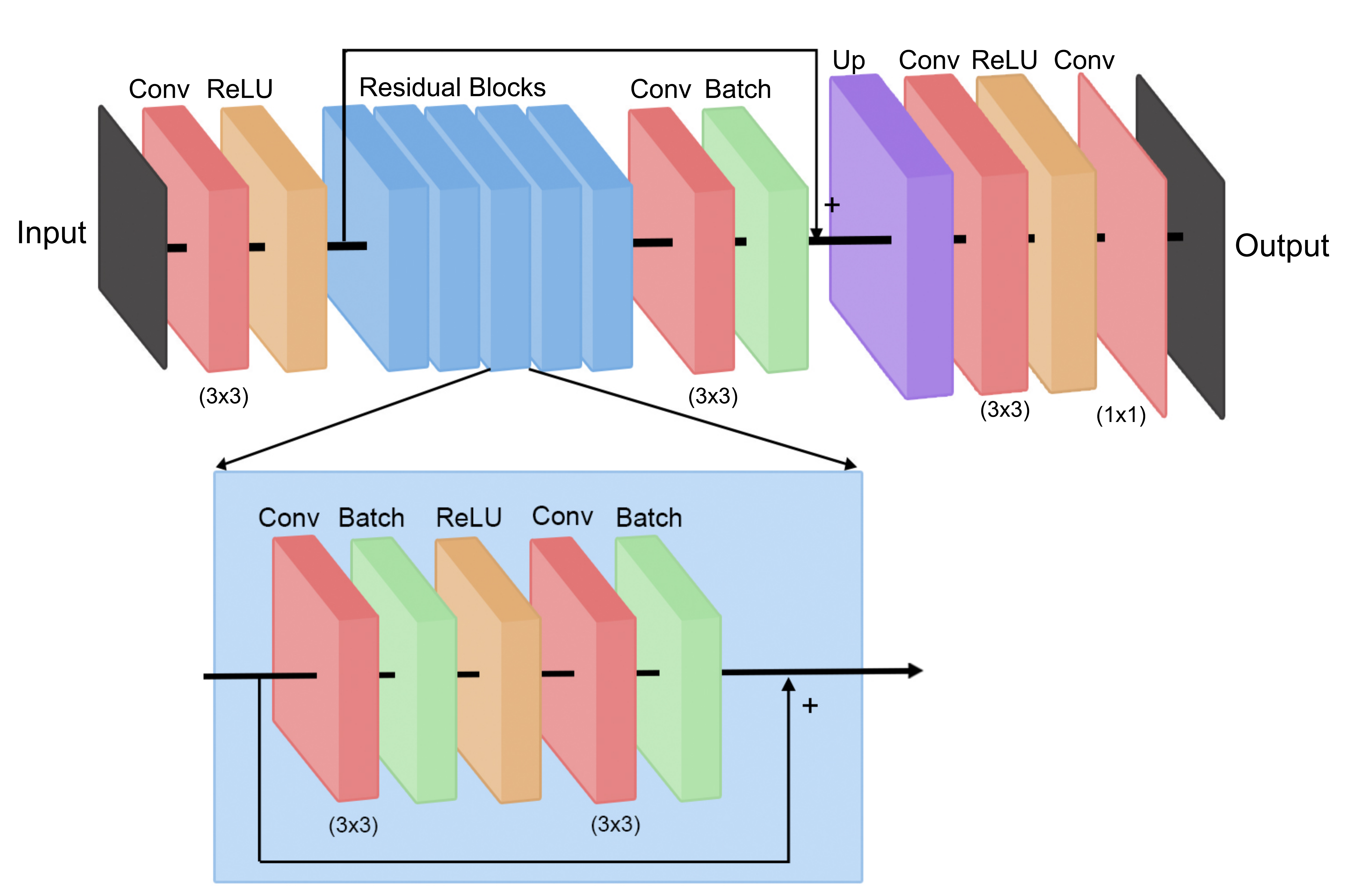}
\caption{Upper panel: architecture of the fully convolutional neural network
used in this work. Colors refer to different types of layers, which are indicated
in the upper labels. The kernel size of convolutional layers are also indicated in 
the lower labels. Black layers are representing the input and output layers. Lower panel: the inner structure of a residual block.}
\label{fig:scheme}
\end{figure*}

\subsection{Our architecture}
We describe in the following the specific architecture of the two deep neural networks 
used to deconvolve and superresolve continuum images and magnetograms. It 
could potentially be possible to use a single network to deconvolve and
superresolve both types of images.
However as each type of data has different well defined properties (like the usual range of values, or the sign of the magnitude) we have decided to use two different neural networks, finding remarkable results.
We refer to the set of two deep neural networks as \enhance. 

The deep neural networks used in this work are inspired
by \texttt{DeepVel} \citep{Asensio2017},
used to infer horizontal velocity fields in the solar photosphere.
Figure \ref{fig:scheme} represents a schematic view of the architecture. It is made
of the concatenation of $N$ residual blocks \citep{residual_network16}. 
Each one is composed of several convolutional layers (two in our case) 
followed by batch normalizations and a ReLU layer for the first
convolutional layer. The internal
structure of a residual block is displayed in the blowup\footnote{We note that we use the non-standard 
implementation of a residual block where the second ReLU activation is removed from the 
reference architecture \citep{residual_network16}, which provides better results according to \texttt{https://github.com/gcr/torch-residual-networks}} of Fig.\ref{fig:scheme}.

Following the typical scheme of a residual block, there is also a shortcut connection 
between the input and the output of the block
\cite[see more information in][]{residual_network16,Asensio2017}, so that the input
is added to the output. Very deep networks usually
saturate during training producing higher errors than shallow networks because
of difficulties during training (also known as the degradation problem).
The fundamental reason is that the gradient of the loss function with
respect to parameters in early layers becomes exponentially small (also known as the vanishing gradient problem). Residual networks help avoid this problem obtaining state-of-the-art results without adding any extra parameter and with
practically the same computational complexity. It is based 
on the idea that if $y=F(x)$ represents the desired effect of the block on the
input $x$, it is much simpler for a network to learn the deviations from the input (or residual mapping) that it can called $R(x)=y-x$ than the full map $F(x)$, so that $y=F(x)=R(x)+x$.



In our case, all convolutions are carried out with kernels of size $3 \times 3$ and
each convolutional layer uses 64 such kernels. Additionally,
as displayed in Fig. \ref{fig:scheme}, we also impose another shortcut connection between the input
to the first residual block and the batch normalization layer 
after the last residual block. We have checked that this slightly
increase the quality of the prediction. Noting that a convolution of an $N \times N$
image with a $3 \times 3$ kernel reduces the size of the output to
$(N-2) \times (N-2)$, we augment the input image with 1 pixel in 
each side using a reflection padding to compensate for this
and maintain the size of the input and output.


Because \enhance\ carries out $\times 2$ superresolution, we need to add an upsampling
layer somewhere in the architecture (displayed in violet in Fig. \ref{fig:scheme}).
One can find in the literature two main options to do the upsampling. 
The first one involves upsampling the image just after the input and let the
rest of convolutional layers do the work. The second
involves doing the upsampling just before the output.
Following \cite{Dong2016}, we prefer the second option because
it provides a much faster network, since the convolutions
are applied to smaller images. 
Moreover, to avoid artifacts in the upsampling\footnote{The checkerboard artifacts are nicely explained in \texttt{https://distill.pub/2016/deconv-checkerboard/}.} 
we have implemented a nearest-neighbor resize followed by convolution instead of a more standard
transpose convolution.

The last layer that carries out a $1 \times 1$ convolution is of extreme
importance in our networks. Given that we use ReLU activation layers throughout the
network, it is only in this very last layer where the output gets its sign using
the weights associated to the layer. This is of no importance for intensity images, but turns 
out to be crucial for the signed magnetic field.

\begin{figure*}
\centering
\includegraphics[width=0.9\textwidth,clip]{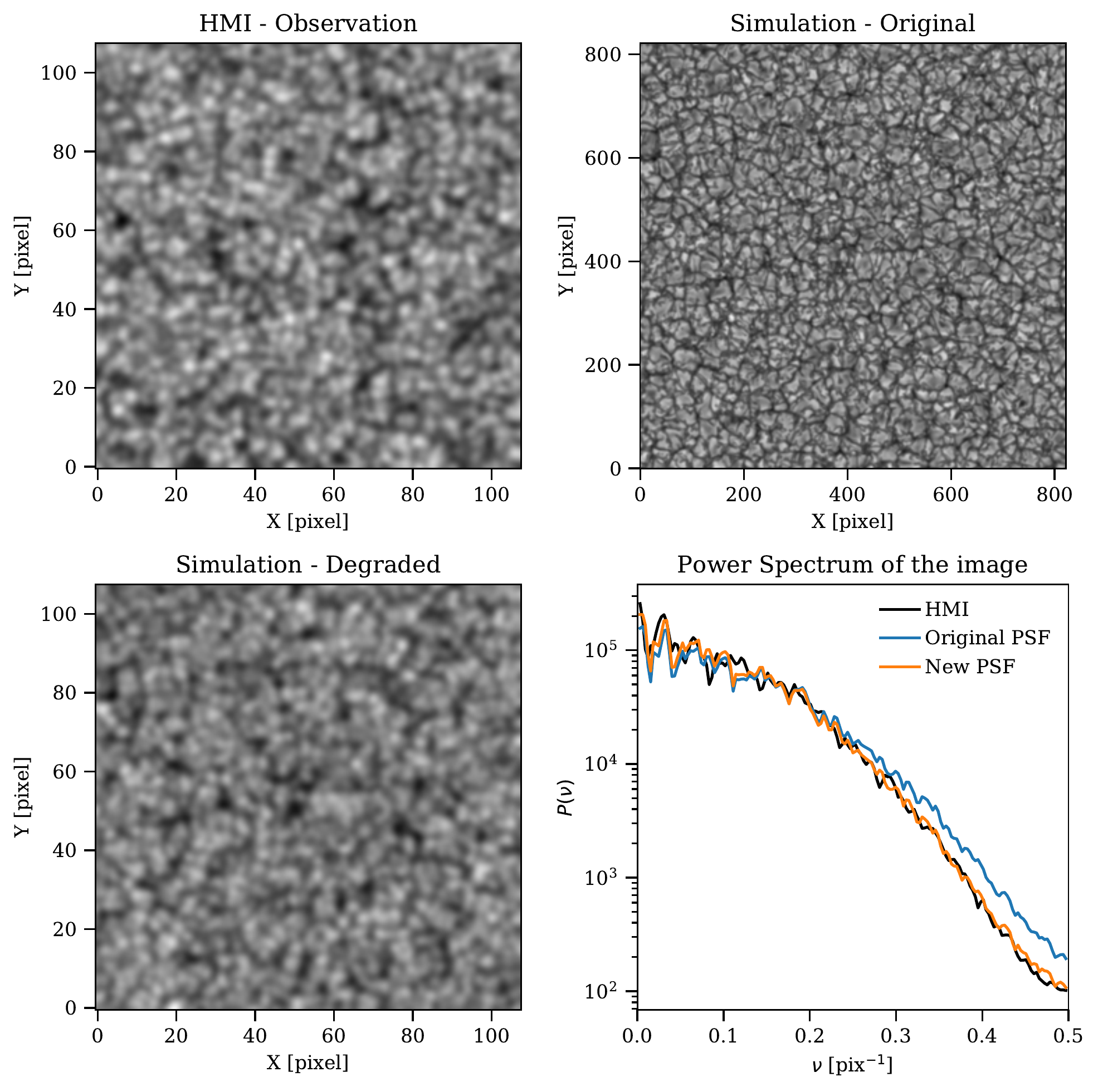}
\caption{Upper left: HMI observation. Upper right: snapshot from the simulation
used for training. Lower left: degraded simulations, which can be compared with
the HMI observations. Lower right: azimuthally averaged power spectrum of
the HMI observations and the degraded simulations with the original PSF and the one modified and used in the training process. The physical dimension of the three maps is 54\arcsec$\times$54\arcsec.}
\label{fig:database}
\end{figure*}

The number of free parameters of our CNN can be easily obtained 
using the previous information. In the scheme of Fig. \ref{fig:scheme}, 
the first  convolution layer generates 64 channels by applying 64 
different kernels of size  $3 \times 3 \times 1$ to the input 
(a single-channel image), using $(3\times3+1)\times 64=640$ 
free parameters. The following convolutional layers have again 
64 kernel filters, but this time each one of size $(3 \times 3 
\times 64 +1)$, with a total of 36928 free parameters. Finally, 
the last layer contains one kernel of size $1 \times 1 \times 64$, 
that computes a weighted average along all channels. The 
total amount of free parameters in this layer is 65 (including the bias).

\subsection{{\bf Our training data and process}}
A crucial ingredient for the success of an CNN is the generation
of a suitable training set of high quality. Our network is 
trained using synthetic continuum images and synthetic
magnetograms from
the simulation of the formation of a solar active region
described by \cite{Cheung2010}. This simulation provides a large
FOV with many solar-like structures (quiet Sun, plage, umbra, penumbra, etc.)
that visually resemble those in the real Sun. We note that if the 
network is trained properly and generalizes well, the network does not
memorize what is in the training set. On the contrary, it applies
what it learns to the new structures. Therefore, we are not
specially concerned by the potential lack of similarity between
the solar structures in the simulation of \cite{Cheung2010}
and the real Sun.

\begin{figure*}
\centering
\includegraphics[width=0.7\linewidth]{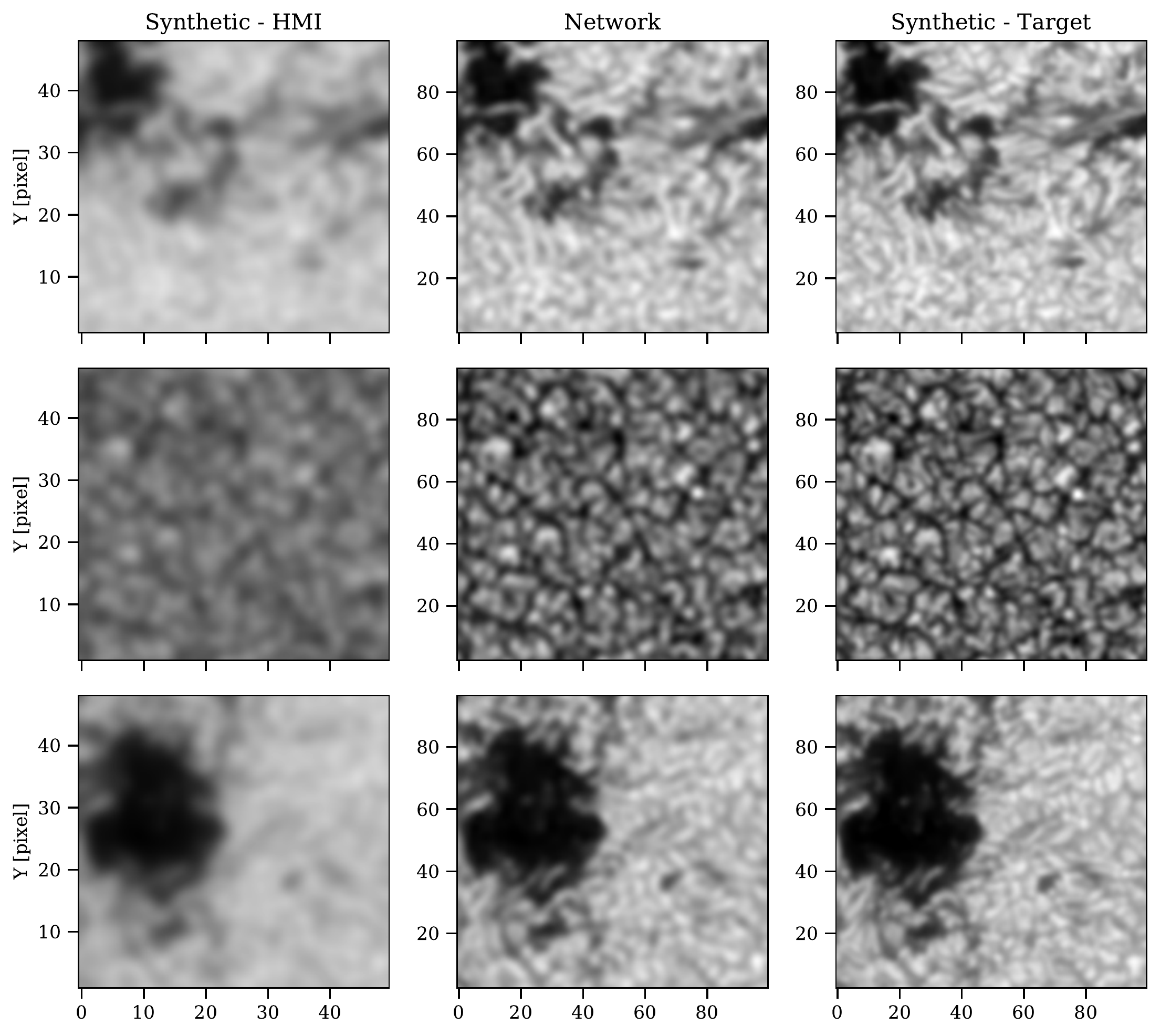}
\includegraphics[width=0.7\linewidth]{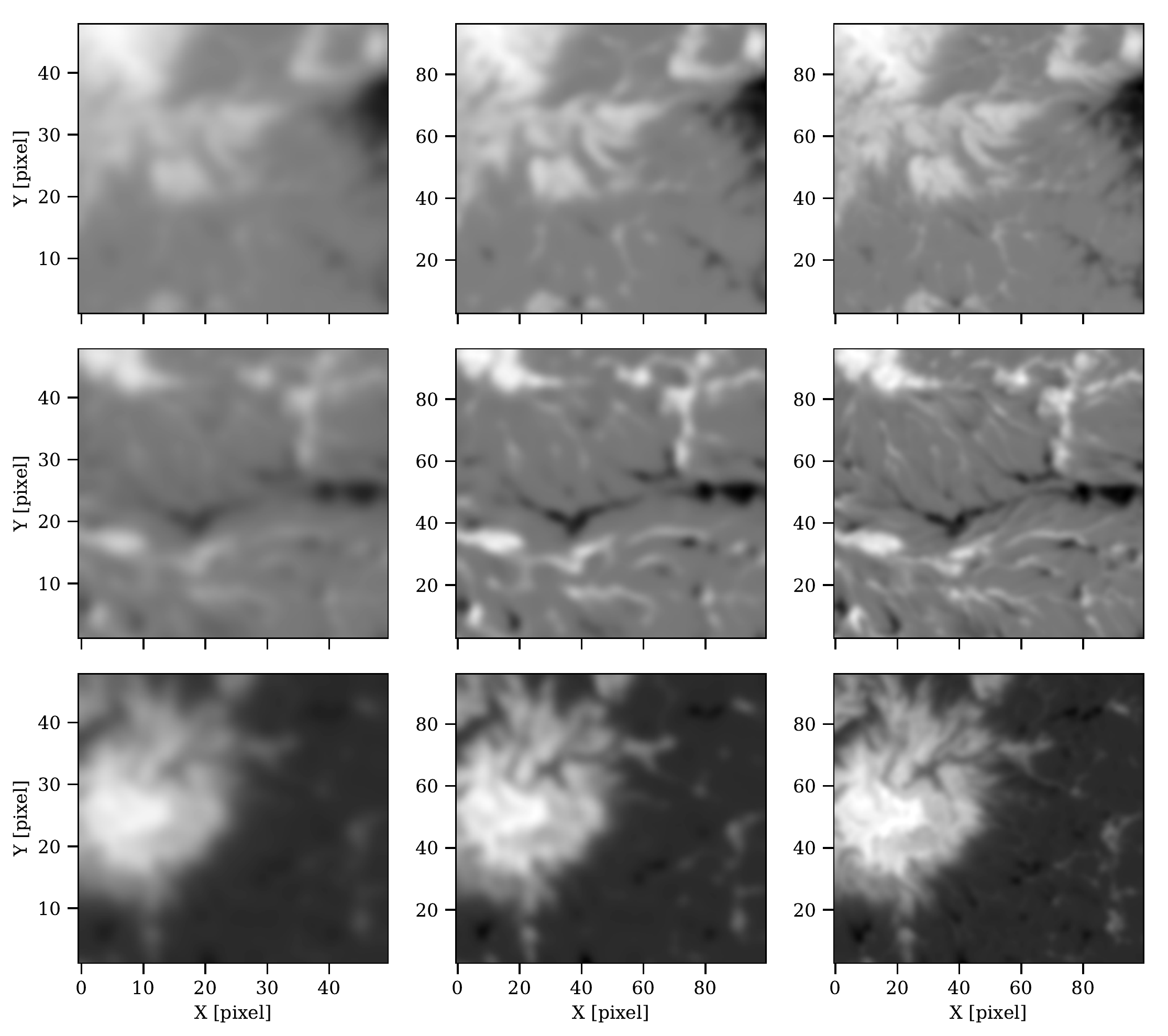}
\caption{Results with the synthetic validation set. The upper three rows
show results for the continuum images, while the lower three rows
display results for the magnetograms. All horizontal and vertical axes
are in pixels.}
\label{fig:validation_synthetic}
\end{figure*}

The radiative MHD simulation was carried out with the
MURaM code \citep{vogler05}. The box spans 92 Mm $\times$
49 Mm in the two horizontal directions and 8.2 Mm in the
vertical direction (with horizontal and vertical grid spacing of
48 and 32 km, respectively). After $\sim$20 h of solar time, an active region
is formed as a consequence of the buoyancy of an injected flux tube in the convection
zone. An umbra, umbral dots, light bridges, and penumbral filaments are
formed during the evolution. As commented before, this constitutes a
very nice dataset of simulated images that look very similar
to those on the Sun. Synthetic gray images are generated from
the simulated snapshots \citep{Cheung2010} and magnetograms
are obtained by just using the vertical magnetic field component
at optical depth unity at 5000~\AA. A total of 250 time
steps are used in the training (slightly less for the magnetograms
when the active region has already emerged to the surface).

We note that the magnetograms of HMI in the \ion{Fe}{i}~6173\,\AA\ correspond to layers in the atmosphere around log$\tau=-1$ \citep{Bello2009}, while our magnetograms are extracted from log$\tau=0$, where $\tau$ is the optical depth at 5000~\AA. In our opinion this will not affect the results
because the concentration of the magnetic field is similar in terms of size and shape in both atmospheric heights.

The synthetic images (and magnetograms) are then treated to simulate a real 
HMI observation. 
All 250 frames of 1920 $\times$ 1024 images are convolved
with the HMI PSF \citep{Wachter2012,Yeo2014,Couvidat2016} and resampled to 0.504\arcsec/pixel. 
For simplicity, we have used the PSF described in \cite{Wachter2012}. The
PSF functional form is azimuthally symmetric and it is given by
\begin{equation}
\mathrm{PSF}(r) = (1-\epsilon) \exp \left[ -\left(\frac{r}{\omega}\right)^2 \right] + 
\epsilon \left[1+\left( \frac{r}{W}\right)^k \right]^{-1},
\end{equation}
which is a linear combination of a Gaussian and a Lorentzian. Note that the 
radial distance is $r=\pi D \theta/\lambda$, with $D$ the telescope
diameter, $\lambda$ the observing wavelength and $\theta$ the distance
in the focal plane in arcsec. The reference values for the 
parameters \citep{Wachter2012} are $\epsilon=0.1$, $\omega=1.8$, $k=3$ and $W=3$. 

Figure \ref{fig:database} demonstrates the similarity between an HMI 
image of the quiet Sun (upper left panel) and the simulations degraded and 
downsampled (lower left panel). The simulation at the original resolution
is displayed in the upper right panel. For clarity, we display the 
horizontal and vertical axis in pixel units, instead of physical units.
This reveals the difference in spatial resolution, both from the
PSF convolution and the resampling. In this process we also realized that using the 
PSF of \cite{Wachter2012}, the azimuthally averaged power spectrum of the degraded 
simulated quiet Sun
turns out to have stronger tails than those of the observation.
For this reason, we slightly modified it so that we finally used
$\omega=2$ and $W=3.4$. The curve with these modified values is 
displayed in orange as the new PSF in 
the Fig. \ref{fig:database} with the original PSF and the default 
values in blue. For consistency, we also applied this
PSF to the magneto-convection simulations described by \cite{stein12_b} and 
\cite{stein12_a}, finding a similar improvement in the comparison
with observations.

One could argue that using the more elaborate PSFs of \cite{Yeo2014} (obtained
via observations of the Venus transit) or \cite{Couvidat2016} (obtained with ground 
data before the launch) is preferred. However, we point out that applying the
PSF of \cite{Wachter2012} (with some modifications that are specified before)
to the simulations produce images that compare excellently at a quantitative
level with the observations. Anyway, given that our code is open sourced, anyone
interested in using a different PSF can easily retrain the deep networks.

Then, we randomly extract 50000 patches of $50\times 50$ pixels both spatially
and temporally, which will constitute the input patches of the
training set. We also randomly extract a smaller subset of 5000 patches
which will act as a validation set to avoid overfitting. These are used 
during the training to check that the CNN generalizes well and is 
not memorizing the training set.
The targets of the training set are obtained similarly but convolving
with the Airy function of a telescope twice the diameter of HMI (28 cm), which gives a 
diffraction limit of $0.55"$/pixel, and then
resampled to $0.25"$/pixel. Therefore, the sizes of the output patches 
are $100 \times 100$ pixels.
All inputs and outputs for the continuum images are normalized to the average 
intensity of the quiet Sun. This is very convenient when the network is 
deployed in production because this quantity $I/Ic$ is almost always available.
On the contrary, the magnetograms are divided by 10$^3$, so they are
treated in kG during the training.

The training of the network is carried out by minimizing a loss function
defined as the squared difference between the output of the network 
and the desired output defined on the training set. To this end, we use 
the Adam stochastic optimizer \citep{adam14} with a learning rate of $\eta=10^{-4}$. 
The training is done in a Titan X GPU for 20 epochs, taking
$\sim 500$ seconds per epoch. We augment the loss function 
with an $\ell_2$ regularization for the 
elements of the kernels of all convolutional layers to avoid overfitting. 
Finally, we add Gaussian noise (with an amplitude of 10$^{-3}$ in units of the 
continuum intensity for
the continuum images and 10$^{-2}$ for the magnetograms, following HMI
standard specifications\footnote{\texttt{http://hmi.stanford.edu/Description/HMI\_Overview.pdf}}) to stabilize
the training and produce better quality predictions. This is important
for regions of low contrast in the continuum images and regions
of weak magnetic fields in the magnetograms. 

Apart from the size and number of kernels, there are a few additional 
hyperparameters that need to be defined in \enhance. 
The most important ones are the number of residual
blocks, the learning rate of the Adam optimizer and the amount
of regularization. We have found
stable training behavior with a learning rate of $10^{-4}$ so 
we have kept this fixed. Additionally, we found
that a regularization weight of $10^{-6}$ for the continuum
images and $10^{-5}$ for the magnetograms provides nice and stable results. 

Finally, five residual blocks with $\sim$450k free parameters provide predictions that are almost
identical to those of 10 and 15 residual blocks but much faster. 
We note that the number of residual blocks can be further decreased even down to one and still
a good behavior is found (even if the number of kernels is decreased to 32). This
version of \enhance\ is 6 times faster than the one presented here, reducing the 
number of parameters to $\sim$40k, with differences
around 3\%. Although \enhance\ is already
very fast, this simplified version can be used for an in-browser online superresolution and
deconvolution of HMI data.

\begin{figure*}
\centering
\includegraphics[height=0.95\textheight]{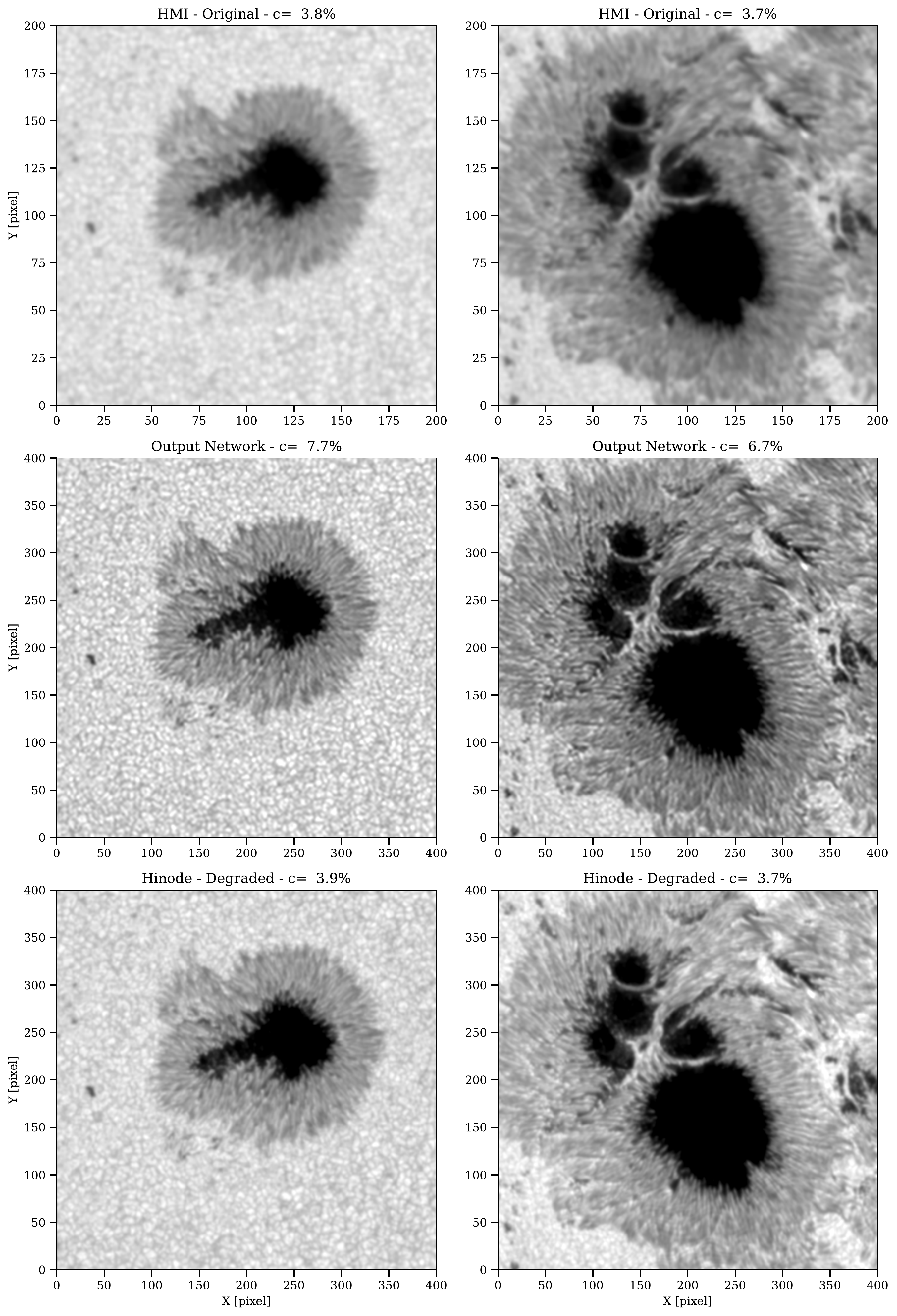}
\caption{This figure shows the application of the neural network to real 
HMI images. From the upper to the lower part of each column: the original 
HMI images, the output of the neural network and the degraded Hinode 
image. All the axis are in pixel units.}
\label{fig:subplot}
\end{figure*}

\begin{figure*}[!ht]
\centering
\includegraphics[width=\linewidth,trim={0 0 0 0},clip]{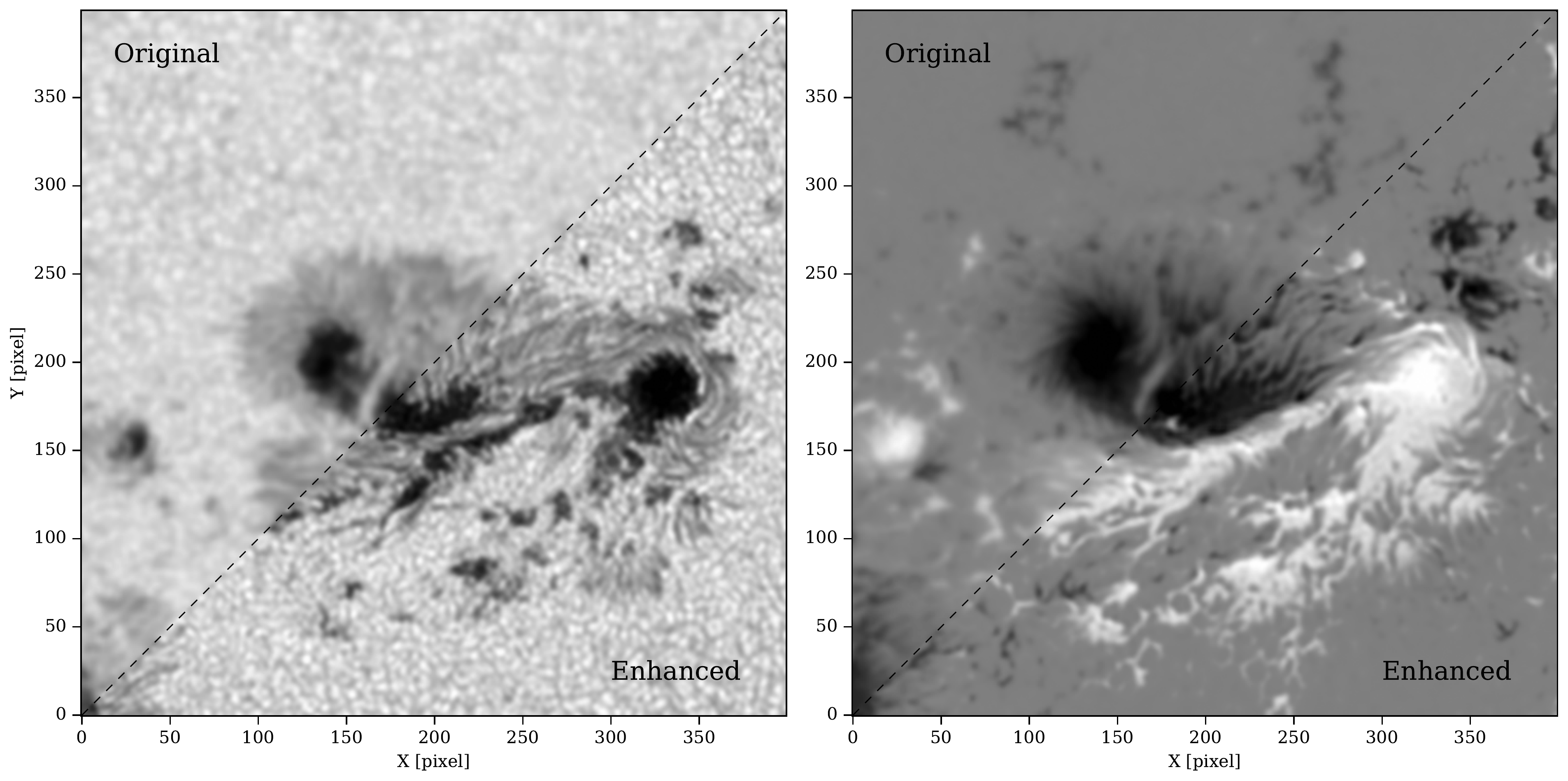}
\caption{This figure show a example our neural network applied to the intensity (left)
and magnetogram (right) for the same region. The FOV is divided 
in two halfs. The upper 
half shows the HMI original image, without applying the neural network. The lower 
half shows enhanced image applying the neural network to the last image. The original image was resampled to have the same scale of the network output.
}
\label{fig:bigplot}
\end{figure*}

\section{Results}
\subsection{Validation with synthetic images}
Before proceeding to applying the networks to real data, we show in Fig.
\ref{fig:validation_synthetic} the results with some of the patches
from the validation set which are not used during the training.
The upper three rows show results for the continuum images, while
the lower three rows show results for the magnetograms. The leftmost column
is the original synthetic image at the resolution of HMI. The rightmost
column is the target that should be recovered by the network, which has
doubled the number of pixels in each dimension. The middle
column displays our single-image superresolution results.

Even though the appearance of all small-scale details are not
exactly similar to the target, we consider that \enhance\ is doing
a very good job in deconvolving and superresolving the data in
the first column. In the regions of increased activity, we find
that we are able to greatly improve the fine structure, specially
in the penumbra. Many details are barely visible in the synthetic HMI
image but can be guessed. Of special relevance are the protrusions
in the umbra in the third row, which are very well recovered by
the neural network.
The network also does a very good job in the quiet Sun, correctly
recovering the expected shape of the granules from the blobby appearance in
the HMI images.

\subsection{In the wild}
The trained networks are then applied to real HMI data. In order to validate 
the output of our neural network we have selected 
observations of the Broadband Filter Instrument (BFI) from the 
Solar Optical Telescope \cite[SOT][]{ichimoto08,tsuneta_hinode08} onboard Hinode \citep{Kosugi2007}.
The pixel size of the BFI is 
$0.109"$ and the selected observations were obtained in the 
red continuum filter at $6684 \pm 2$ \AA, which is the one closer to the
observing wavelength of HMI.
To properly compare our results with Hinode, we have convolved the BFI images with an 
Airy function of a telescope of 28 cm diameter and resampled to $0.25"$/pixel 
to match those of the output of \enhance. 
The Hinode images have not been deconvolved from the influence
of its PSF. We point out that the long tails of the PSF of the Hinode/SOT instrument produces a
slight decrease of the contrast \citep{Danilovic2010} and this is the reason why
our enhanced images have a larger contrast. 


\begin{figure*}[!ht]
\centering
\includegraphics[width=0.495\linewidth]{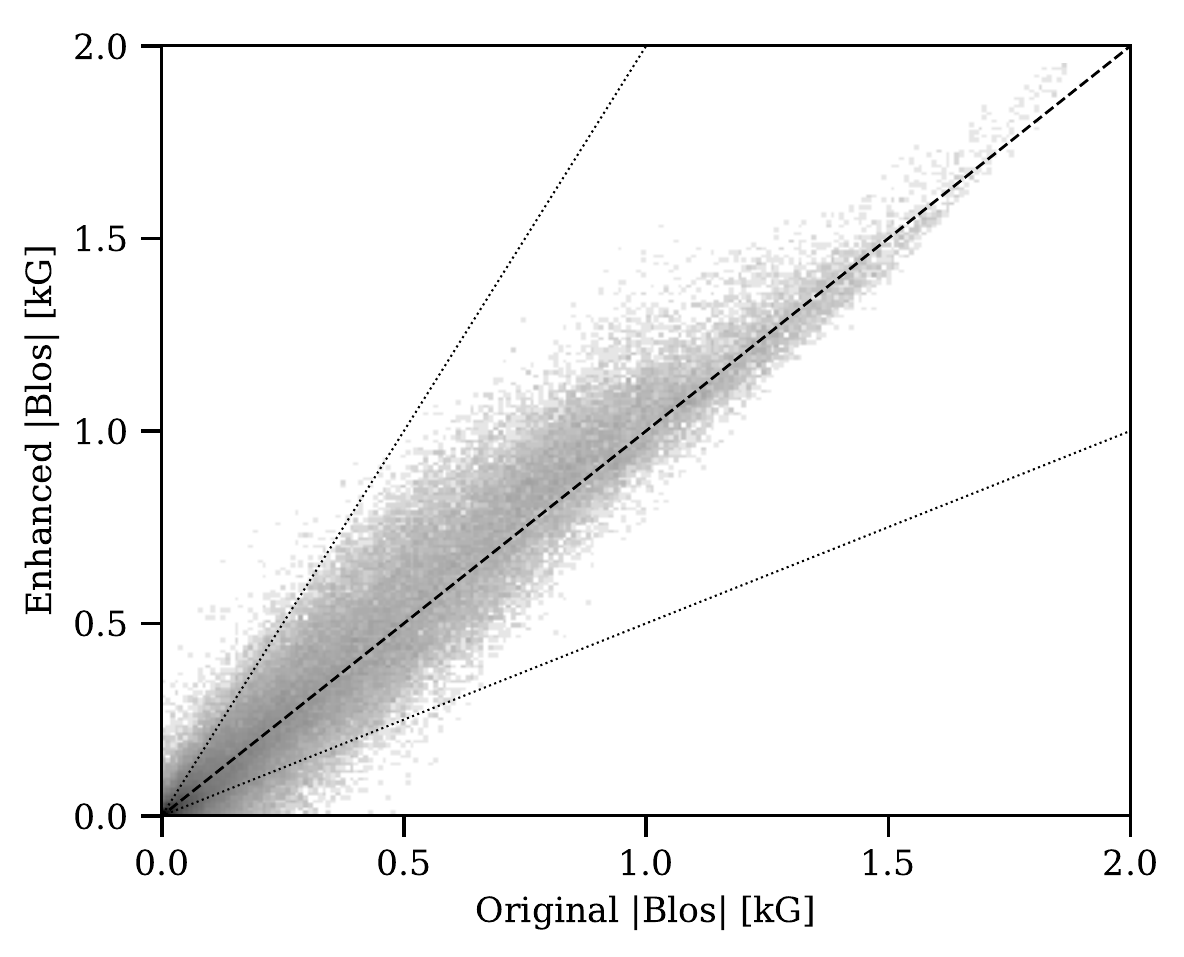}
\includegraphics[width=0.495\linewidth]{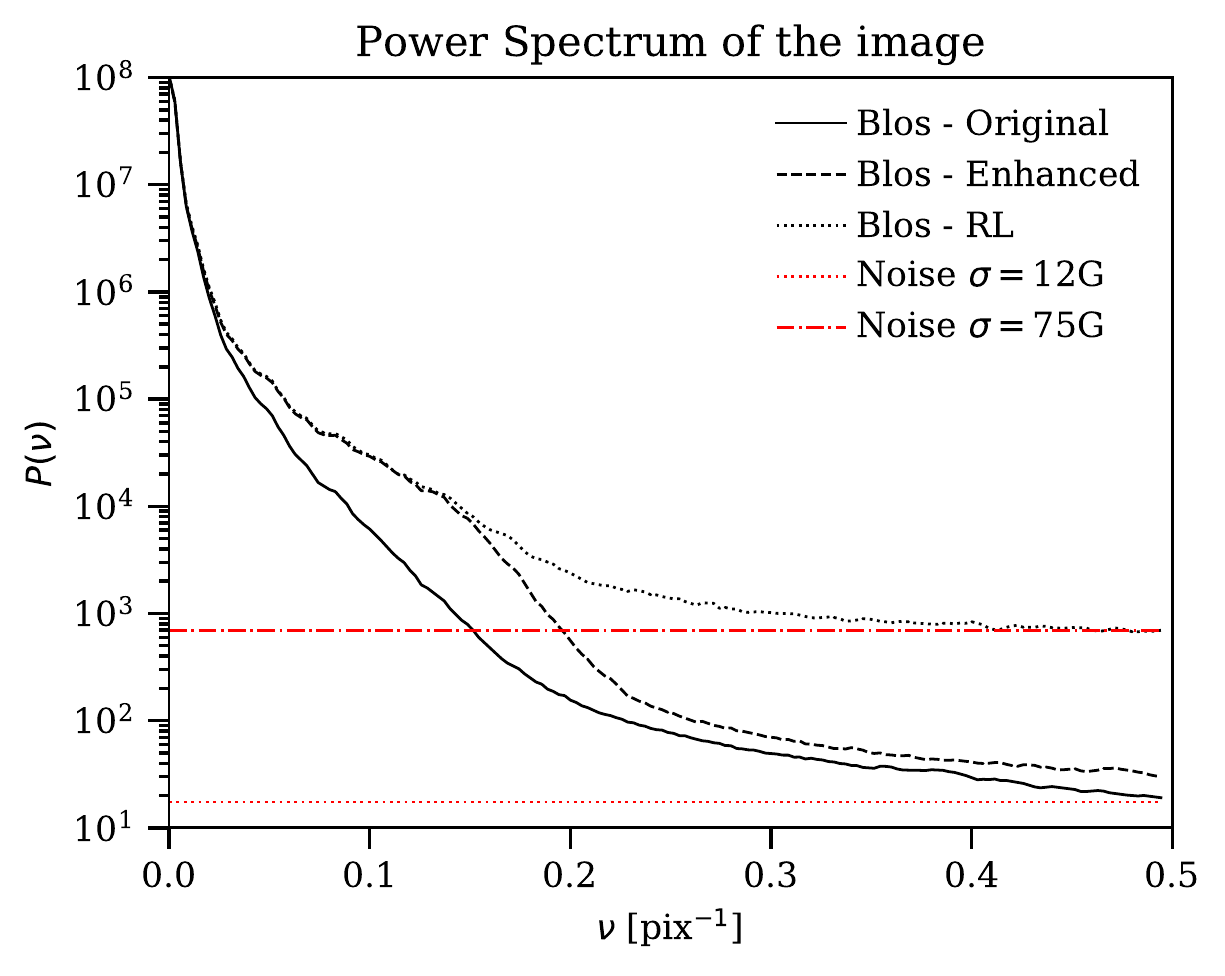}
\caption{Left: scatter plot between the original magnetogram signal 
and the deconvolved magnetogram signal. Dotted lines indicate a change of a factor 2. 
Right: spatial Fourier power spectrum of all considered 
magnetograms: the original, the output of \enhance\ and the one deconvolved with RL. We
also show the power spectrum of white noise at two different levels.}
\label{fig:powerBlos}
\end{figure*}

\subsubsection{Continuum images}

Figure \ref{fig:subplot}
displays this comparison for two different regions (columns) observed simultaneously
with Hinode and HMI. 
These two active regions are: NOAA 11330 (N09, E04) observed on October 27, 2011 
(first column) and NOAA 12192 (S14, E05) observed on October 22, 2014 (second column).
We have used HMI images with a cadence of 45 seconds, which is the
worst scenario in terms of noise in the image. 
The upper rows show the original HMI images. The 
lower rows display the degraded Hinode images, while the central row shows
the output of our neural network. Given the fully convolutional character of the 
deep neural network used in this work, it can be applied seamlessly to input 
images of arbitrary size. As an example, an image of size $400 \times 400$
can be superresolved and deconvolved in $\sim$100 ms using a Titan X GPU, or $\sim$1 s using a 3.4 GHz Intel Core i7.

The contrast $\sigma_I/I$, calculated as the standard
deviation of the continuum intensity divided by the
average intensity of the area, is quoted in the title of each
panel and has been obtained in a small region of the image 
displaying only granulation.
The granulation contrast increases from $\sim3.7$\% to $\sim$7\% \citep[as][]{Couvidat2016},
almost a factor 2 larger than the one provided by degraded Hinode. 
Note that the contrast may be slightly off for the right column because of the small
quiet Sun area available. The granulation contrast measured in 
Hinode without degradation is around 7\%. After the resampling, it goes
down to the values quoted in the figure. We note that \citep{Danilovic2008}
analyzed the Hinode granulation contrast at 630 nm and concluded that it is 
consistent with those predicted by the simulations (in the range 14$-$15\%)
once the PSF is taken into account.
Just from the visual point of view, it is clear that \enhance\
produces small-scale structures that are almost absent in the HMI images
but clearly present in the Hinode images. Additionally, the deconvolved
and superresolved umbra intensity decreases between 3 and 7\% when compared
to the original HMI umbral intensity.

Interesting cases are the large light bridge in the images of the right 
column, that increases in spatial complexity. Another examples are the 
regions around the light
bridge, that are plagued with small weak umbral dots that are evident
in Hinode data but completely smeared out in HMI. For instance, 
the region connecting the light bridge at $(125,240)$ with the penumbra. Another
similar instance of this enhancement occurs $(375,190)$, a pore with
some umbral dots that are almost absent in the HMI images.

As a caveat, we warn the users that the predictions of the neural network
in areas close to the limb is poorer than those at disk center. Given
that \enhance\ was trained with images close to disk center, one could
be tempted to think that a lack of generalization is the cause for the failure. However,
we note that structures seen in the limb like elongated granules share some similarity
to some penumbral filaments, so these cases are already present in the training set.
The fundamental reason for the failure is that the spatial contrast in the limb 
is very small so the neural
network does not know how to reconstruct the structures, thus creating artifacts. We
speculate that these artifacts will not be significantly reduced even if limb
synthetic observations are included in the training set.

\subsubsection{A magnetogram example: AR 11158}
As a final example, we show in Fig. \ref{fig:bigplot} an example of 
the neural network applied to the intensity and the magnetogram for the 
same region: the NOAA 11158 (S21, W28), observed on February 15, 2011.
The FOV is divided in two halfs. The upper parts show the HMI original 
image both for the continuum image (left panel) and the magnetogram (right panel). 
The lower parts display the enhanced images after applying the neural network.

After the deconvolution of the magnetogram, we find: i) regions with very nearby opposite
polarities suffer from an apparent cancellation in HMI data that can be
restored with \enhance, giving rise to an increase in the 
absolute value of the longitudinal field; and ii) regions far from 
magnetized areas do get contaminated by the surroundings in HMI, which
are also compensated for with \enhance, returning smaller longitudinal
fields. The left panel of Fig. \ref{fig:powerBlos} shows the density
plot of the input vs. output longitudinal magnetic field. Almost all the 
points lie in the 1:1 relation. However, points around 1 kG for HMI are
promoted to larger values in absolute value, a factor $\sim 1.3 - 1.4$ higher 
than the original image \citep{Couvidat2016}. 

Another interesting point to study is the range of spatial scales at which
\enhance\ is adding information. The right panel of 
Fig. \ref{fig:powerBlos} displays the 
power spectrum of both magnetograms showed in the right part of 
Fig. \ref{fig:bigplot}. The main difference between both curves is 
situated in the range of spatial scales $\nu = 0.05-0.25$ pix$^{-1}$ with a peak 
at $\nu=0.15$ pix$^{-1}$. In other words, the neural network is operating 
mainly at scales between 4 and 20 pixels, where the smearing effect of the PSF 
is higher.

The same effect can be seen when a standard Richardson--Lucy 
maximum-likelihood algorithm (RL) (including a bilinear interpolation to
carry out the superresolution)
is used (see Section \ref{sec:RL} for more details). The power spectrum of the output 
of \enhance\ and the one deconvolved with RL 
are almost the same for frequencies below 0.15 pix$^{-1}$ (equivalent to scales above
$\sim 6$ pix). For larger frequencies (smaller scales), the RL version adds noisy 
small scale structures at a level of $\sim$80 G, that is not case with \enhance. We note that 
the original image has a noise around $\sim$10 G. To quantify this last point, we have 
showed in Fig. \ref{fig:powerBlos} the flat spectrum of white noise artificial images with zero mean and standard deviations $\sigma=12$G and $\sigma=75$G.

\begin{figure*}[!ht]
\centering
\includegraphics[width=\linewidth]{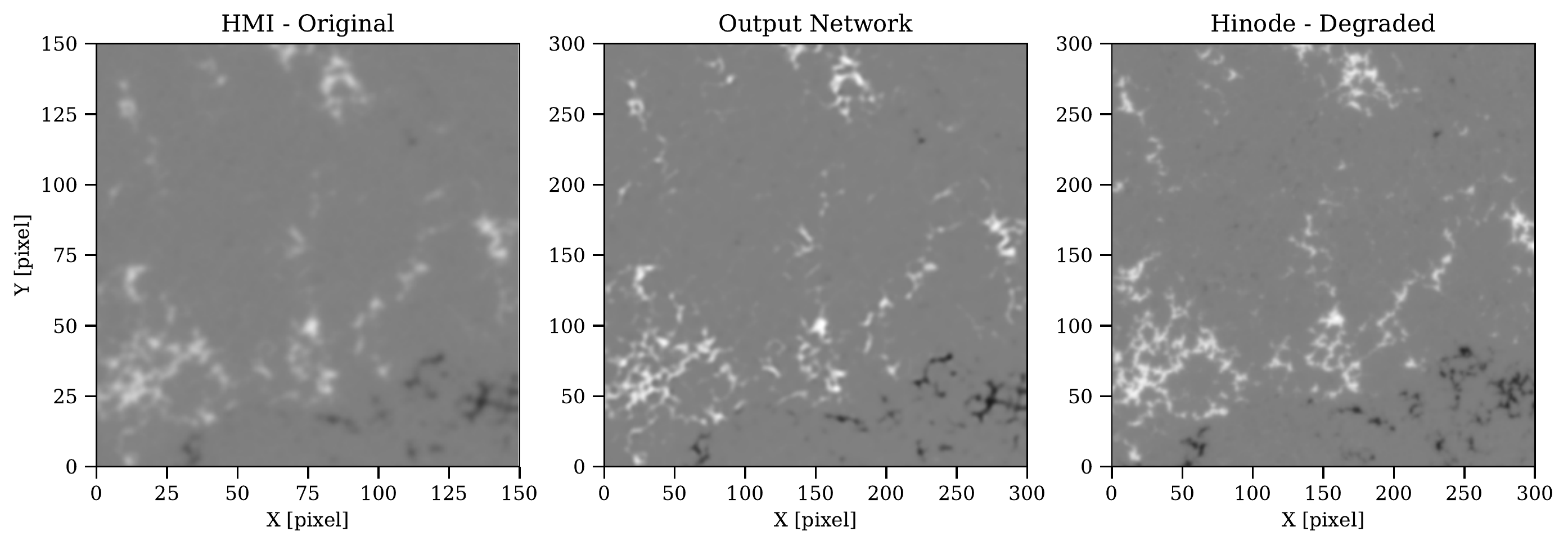}
\caption{Left: original HMI magnetogram of 
a plage region observed on April 25, 2015. Middle: the result of applying \enhance\ to 
the HMI magnetogram. Right: the Hinode magnetogram at the same resolution of \enhance. The magnetic flux has been clipped from $-$1kG to 1kG.}
\label{fig:compararPlage}
\end{figure*}

\subsubsection{Other general properties}

Depending on the type of structure analyzed, the effect of the deconvolution is 
different. In plage regions, where the magnetic areas are less clustered than in
a sunspot, the impact of the stray light is higher. Then, \enhance\ produces a magnetic 
field that can increase up to a factor 2 \citep{Yeo2014}, with magnetic structures smaller in
size, as signal smeared onto the surrounding quiet Sun 
is put back on its original location. According to the left panel of Fig \ref{fig:powerBlos}, fields
with smaller amplitudes suffer a larger relative change. As a guide to the eye, the two dotted 
lines indicate a change of a factor 2 in the same figure.

To check these conclusions, we have used a Hinode-SOT 
Spectropolarimeter (SP) \citep{Lites2013} Level 1D\footnote{\texttt{http://sot.lmsal.com/data/sot/level1d/}}
magnetogram. The region was observed 
in April 25, 2015 at 04:00h UT and its pixel size is around 0.30''/pix.  
Figure \ref{fig:compararPlage} shows the increase of the magnetic field after the 
deconvolution: magnetic fields of kG flux were diluted by the PSF and recovered with \enhance. 
It was impossible to find the Hinode map of exactly the same region at exactly the same moment, 
so that some differences are visible. However the general details are retrieved. 
In regions of strong concentrations, like the ones found in Fig. \ref{fig:bigplot}, 
almost each polarity is spatially concentrated and increased by a factor below 1.5.

The magnetogram case is more complex than the intensity map. Many studies 
\citep{Krivova2004,Pietarila2013,Bamba2014} have demonstrated the influence of 
the resolution to estimate a value of the magnetic flux and products of magnetogram 
as nonlinear force-free extrapolations \citep{Tadesse2013,DeRosa2015}, to compare
with in--situ spacecraft measurements \citep{Linker2017}.

Contrary to deconvolving intensity images, deconvolving 
magnetograms is always a very delicate issue. 
The difficulty relies on the presence of cancellation produced during 
the smearing with a PSF if magnetic elements 
of opposite polarities are located nearby. This never happens for 
intensity images, which are always non-negative. Consequently, one can arbitrarily increase
the value of nearby positive and negative polarities while maintaining a small
quadratic approximation to the desired output.
This effect is typically seen when a standard RL algorithm is used for
deconvolution.

\enhance\ avoids this effect by learning suitable spatial priors from 
the training dataset. It is true that the method will not be able
to separate back two very nearby opposite polarities that have been
fully canceled by the smearing of the PSF. Extensive tests show that
the total absolute flux of each deconvolved image is almost the same as that
in the original image, i.e., the magnetic 
field is mainly "reallocated".


\subsubsection{Comparison with a standard RL deconvolution algorithm}
\label{sec:RL}

\begin{figure}
\centering
\includegraphics[width=0.45\textwidth]{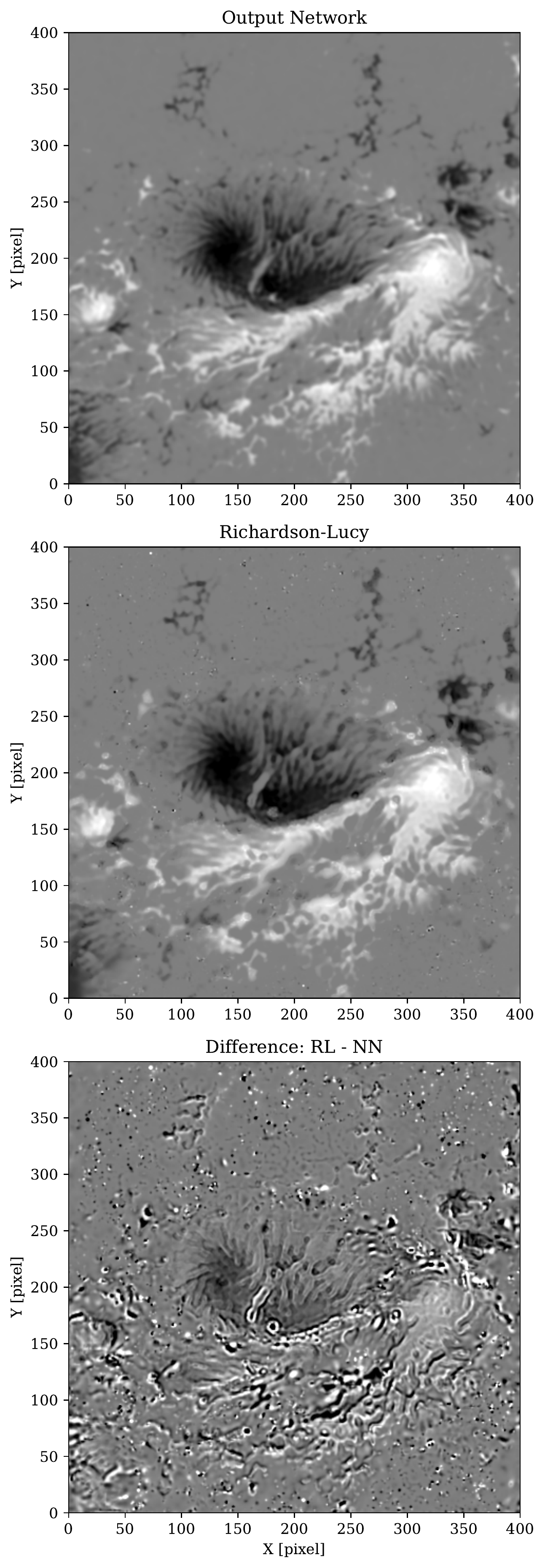}
\caption{ Upper: the output of \enhance. Middle: output after applying a Richardson-Lucy method to deconvolve the image. Lower: the difference between the RL version and the \enhance\ output. The magnetic flux has been clipped to $\pm$1.5kG and $\pm$100G in the last image.}
\label{fig:compara2}
\end{figure}

As a final step, we compare our results with those of a RL algorithm in a
complicated case. Fig. \ref{fig:compara2} shows the same image deconvolved with 
both methods. The output of \enhance\ 
is similar to the output of the RL method. Some noisy artifacts are detected in areas 
with low magnetic field strength.

A detailed analysis can reveal some differences, tough. 
In the light--bridge (LB), the magnetic field is lower in the RL version. 
Additionally, the polarity inversion line (PIL) appears more enhanced and splitted 
in the RL version than in the \enhance\ one. The magnetic flux in both areas 
(LB and PIL) are reduced by a factor 0.5, which might be an indication of
too many iterations. The magnetic field 
strength of the umbra is between 50~G and 80~G higher in the RL version.

As a final test, we have checked the difference between the original image and the 
output of \enhance\ convolved with the PSF. The average relative difference is around 
4\% (which is in the range 10-80~G depending on the flux of the pixel), 
which goes down to less than 1\% in the RL case (this is a clear indication
that \enhance\ is introducing prior information not present in the data). Additionally,
our network is orders of magnitude faster than RL, it does not create 
noisy artifacts and the estimation of the magnetic field is as robust 
as a RL method.

\section{Conclusions and future work}
This paper presents the first successful deconvolution and super-resolution
applied on solar images using deep convolutional neural network. 
It represents, after \cite{Asensio2017}, a new step
toward the implementation of new machine learning techniques in the field
of Solar Physics.

Single-image superresolution and deconvolution, either for continuum images or for
magnetograms, is an ill-defined problem. It requires the addition
of extra knowledge for what to expect in the high-resolution
images. The deep learning approach presented in this paper extracts this knowledge from
the simulations and also applies a deconvolution. All this
is done very fast, almost in real-time, and to images of arbitrary size.
We hope that \enhance\ will allow researchers 
to study small-scale details in HMI images and magnetograms, something
that cannot be currently done.

Often, HMI is used not as the primary source of information but
as a complement for ground-based observations, providing the context. 
For this reason, having enhanced images where you can analyze the context
with increased resolution is interesting.

We have preferred to be conservative and only do superresolution by a factor 2. We have
carried out some tests with a larger factor, but the results were not satisfactory. It remains
to test whether other techniques proposed in this explosively growing field can work 
better. Among others, techniques like a gradual up-sampling  \citep{Zhao2017}, recursive convolutional 
layers \citep{Kim2015}, recursive residual blocks \citep{Ying2017} or using adversarial
networks as a more elaborate loss function \citep{Ledig2016,schawinski17} can
potentially produce better results.

We open-source \enhance\footnote{\texttt{https://github.com/cdiazbas/enhance}}, providing the 
methods to apply the trained networks used in this work to HMI images
or re-train them using new data. In the future, we plan to extend the technique to other telescopes/instruments to generate superresolved and deconvolved images.

\begin{acknowledgements}
We would like to thank the anonymous referee for its comments and suggestions.
We thank Mark Cheung for kindly sharing with us the simulation data, without which this study 
would not have been possible. 
Financial support by the Spanish Ministry of Economy and Competitiveness
through project AYA2014-60476-P is gratefully acknowledged.
CJDB acknowledges Fundaci\'on La Caixa for the financial support 
received in the form of a PhD contract.
We also thank the NVIDIA Corporation for the donation of the Titan X GPU 
used in this research.
This research has made use of NASA's Astrophysics Data System 
Bibliographic Services.
We acknowledge the community effort devoted to the development of the following 
open-source packages that were used in this work: 
\texttt{numpy} (\texttt{numpy.org}), \texttt{matplotlib} (\texttt{matplotlib.org}),
\texttt{Keras} (\texttt{keras.io}), and \texttt{Tensorflow} {\texttt{tensorflow.org}}
and \texttt{SunPy} {\texttt{sunpy.org}}.
\end{acknowledgements}







\bibliographystyle{mnras}
\bibliography{aanda.bbl}

\begin{thebibliography}{56}
\expandafter\ifx\csname natexlab\endcsname\relax\def\natexlab#1{#1}\fi

\bibitem[{{Asensio Ramos} \& {de la Cruz Rodr{\'{\i}}guez}(2015)}]{Asensio2015}
{Asensio Ramos}, A. \& {de la Cruz Rodr{\'{\i}}guez}, J. 2015, \aap, 577, A140

\bibitem[{{Asensio Ramos} {et~al.}(2017){Asensio Ramos}, {Requerey}, \&
  {Vitas}}]{Asensio2017}
{Asensio Ramos}, A., {Requerey}, I.~S., \& {Vitas}, N. 2017, \aap, 604, A11

\bibitem[{{Asensio Ramos} \& {Socas-Navarro}(2005)}]{asensio05}
{Asensio Ramos}, A. \& {Socas-Navarro}, H. 2005, \aap, 438, 1021

\bibitem[{{Bamba} {et~al.}(2014){Bamba}, {Kusano}, {Imada}, \&
  {Iida}}]{Bamba2014}
{Bamba}, Y., {Kusano}, K., {Imada}, S., \& {Iida}, Y. 2014, \pasj, 66, S16

\bibitem[{{Bello Gonz{\'a}lez} {et~al.}(2009){Bello Gonz{\'a}lez}, {Yelles
  Chaouche}, {Okunev}, \& {Kneer}}]{Bello2009}
{Bello Gonz{\'a}lez}, N., {Yelles Chaouche}, L., {Okunev}, O., \& {Kneer}, F.
  2009, \aap, 494, 1091

\bibitem[{Bishop(1996)}]{B96}
Bishop, C.~M. 1996, Neural networks for pattern recognition (Oxford University
  Press)

\bibitem[{Borman \& Stevenson(1998)}]{Borman1998}
Borman, S. \& Stevenson, R.~L. 1998, Midwest Symposium on Circuits and Systems,
  374

\bibitem[{{Carroll} \& {Kopf}(2008)}]{carroll08}
{Carroll}, T.~A. \& {Kopf}, M. 2008, \aap, 481, L37

\bibitem[{{Cheung} {et~al.}(2010){Cheung}, {Rempel}, {Title}, \&
  {Sch{\"u}ssler}}]{Cheung2010}
{Cheung}, M.~C.~M., {Rempel}, M., {Title}, A.~M., \& {Sch{\"u}ssler}, M. 2010,
  \apj, 720, 233

\bibitem[{{Ciuca} {et~al.}(2017){Ciuca}, {Hern{\'a}ndez}, \&
  {Wolman}}]{Ciuca17}
{Ciuca}, R., {Hern{\'a}ndez}, O.~F., \& {Wolman}, M. 2017, ArXiv e-prints
  [\eprint[arXiv]{1708.08878}]

\bibitem[{{Colak} \& {Qahwaji}(2008)}]{colak08}
{Colak}, T. \& {Qahwaji}, R. 2008, \solphys, 248, 277

\bibitem[{{Couvidat} {et~al.}(2016){Couvidat}, {Schou}, {Hoeksema}, {Bogart},
  {Bush}, {Duvall}, {Liu}, {Norton}, \& {Scherrer}}]{Couvidat2016}
{Couvidat}, S., {Schou}, J., {Hoeksema}, J.~T., {et~al.} 2016, \solphys, 291,
  1887

\bibitem[{{Danilovic} {et~al.}(2008){Danilovic}, {Gandorfer}, {Lagg},
  {Sch{\"u}ssler}, {Solanki}, {V{\"o}gler}, {Katsukawa}, \&
  {Tsuneta}}]{Danilovic2008}
{Danilovic}, S., {Gandorfer}, A., {Lagg}, A., {et~al.} 2008, \aap, 484, L17

\bibitem[{{Danilovic} {et~al.}(2010){Danilovic}, {Sch{\"u}ssler}, \&
  {Solanki}}]{Danilovic2010}
{Danilovic}, S., {Sch{\"u}ssler}, M., \& {Solanki}, S.~K. 2010, \aap, 513, A1

\bibitem[{{DeRosa} {et~al.}(2015){DeRosa}, {Wheatland}, {Leka}, {Barnes},
  {Amari}, {Canou}, {Gilchrist}, {Thalmann}, {Valori}, {Wiegelmann},
  {Schrijver}, {Malanushenko}, {Sun}, \& {R{\'e}gnier}}]{DeRosa2015}
{DeRosa}, M.~L., {Wheatland}, M.~S., {Leka}, K.~D., {et~al.} 2015, \apj, 811,
  107

\bibitem[{{Dong} {et~al.}(2015){Dong}, {Change Loy}, {He}, \&
  {Tang}}]{Dong2015}
{Dong}, C., {Change Loy}, C., {He}, K., \& {Tang}, X. 2015, ArXiv e-prints
  [\eprint[arXiv]{1501.00092}]

\bibitem[{{Dong} {et~al.}(2016){Dong}, {Change Loy}, \& {Tang}}]{Dong2016}
{Dong}, C., {Change Loy}, C., \& {Tang}, X. 2016, ArXiv e-prints
  [\eprint[arXiv]{1608.00367}]

\bibitem[{{Hayat}(2017)}]{Hayat2017}
{Hayat}, K. 2017, ArXiv e-prints [\eprint[arXiv]{1706.09077}]

\bibitem[{{He} {et~al.}(2015){He}, {Zhang}, {Ren}, \&
  {Sun}}]{residual_network16}
{He}, K., {Zhang}, X., {Ren}, S., \& {Sun}, J. 2015, ArXiv e-prints
  [\eprint[arXiv]{1512.03385}]

\bibitem[{{Ichimoto} {et~al.}(2008){Ichimoto}, {Lites}, {Elmore}, {Suematsu},
  {Tsuneta}, {Katsukawa}, {Shimizu}, {Shine}, {Tarbell}, {Title}, {Kiyohara},
  {Shinoda}, {Card}, {Lecinski}, {Streander}, {Nakagiri}, {Miyashita},
  {Noguchi}, {Hoffmann}, \& {Cruz}}]{ichimoto08}
{Ichimoto}, K., {Lites}, B., {Elmore}, D., {et~al.} 2008, \solphys, 249, 233

\bibitem[{Ioffe \& Szegedy(2015)}]{batch_normalization15}
Ioffe, S. \& Szegedy, C. 2015, in Proceedings of the 32nd International
  Conference on Machine Learning (ICML-15), ed. D.~Blei \& F.~Bach (JMLR
  Workshop and Conference Proceedings), 448--456

\bibitem[{{Kim} {et~al.}(2015){Kim}, {Lee}, \& {Lee}}]{Kim2015}
{Kim}, J., {Lee}, J.~K., \& {Lee}, K.~M. 2015, ArXiv e-prints
  [\eprint[arXiv]{1511.04491}]

\bibitem[{{Kingma} \& {Ba}(2014)}]{adam14}
{Kingma}, D.~P. \& {Ba}, J. 2014, ArXiv e-prints [\eprint[arXiv]{1412.6980}]

\bibitem[{{Kosugi} {et~al.}(2007){Kosugi}, {Matsuzaki}, {Sakao}, {Shimizu},
  {Sone}, {Tachikawa}, {Hashimoto}, {Minesugi}, {Ohnishi}, {Yamada}, {Tsuneta},
  {Hara}, {Ichimoto}, {Suematsu}, {Shimojo}, {Watanabe}, {Shimada}, {Davis},
  {Hill}, {Owens}, {Title}, {Culhane}, {Harra}, {Doschek}, \&
  {Golub}}]{Kosugi2007}
{Kosugi}, T., {Matsuzaki}, K., {Sakao}, T., {et~al.} 2007, \solphys, 243, 3

\bibitem[{{Krivova} \& {Solanki}(2004)}]{Krivova2004}
{Krivova}, N.~A. \& {Solanki}, S.~K. 2004, \aap, 417, 1125

\bibitem[{LeCun \& Bengio(1998)}]{LeCun1998}
LeCun, Y. \& Bengio, Y. 1998, in The Handbook of Brain Theory and Neural
  Networks, ed. M.~A. Arbib (Cambridge, MA, USA: MIT Press), 255--258

\bibitem[{LeCun {et~al.}(1998)LeCun, Bottou, Orr, \& M\"{u}ller}]{LeCun1998b}
LeCun, Y., Bottou, L., Orr, G.~B., \& M\"{u}ller, K.-R. 1998, in Neural
  Networks: Tricks of the Trade, This Book is an Outgrowth of a 1996 NIPS
  Workshop (London, UK, UK: Springer-Verlag), 9--50

\bibitem[{{Ledig} {et~al.}(2016){Ledig}, {Theis}, {Huszar}, {Caballero},
  {Cunningham}, {Acosta}, {Aitken}, {Tejani}, {Totz}, {Wang}, \&
  {Shi}}]{Ledig2016}
{Ledig}, C., {Theis}, L., {Huszar}, F., {et~al.} 2016, ArXiv e-prints
  [\eprint[arXiv]{1609.04802}]

\bibitem[{{Linker} {et~al.}(2017){Linker}, {Caplan}, {Downs}, {Riley}, {Mikic},
  {Lionello}, {Henney}, {Arge}, {Liu}, {Derosa}, {Yeates}, \&
  {Owens}}]{Linker2017}
{Linker}, J.~A., {Caplan}, R.~M., {Downs}, C., {et~al.} 2017, ArXiv e-prints
  [\eprint[arXiv]{1708.02342}]

\bibitem[{{Lites} {et~al.}(2013){Lites}, {Akin}, {Card}, {Cruz}, {Duncan},
  {Edwards}, {Elmore}, {Hoffmann}, {Katsukawa}, {Katz}, {Kubo}, {Ichimoto},
  {Shimizu}, {Shine}, {Streander}, {Suematsu}, {Tarbell}, {Title}, \&
  {Tsuneta}}]{Lites2013}
{Lites}, B.~W., {Akin}, D.~L., {Card}, G., {et~al.} 2013, \solphys, 283, 579

\bibitem[{Nair \& Hinton(2010)}]{relu10}
Nair, V. \& Hinton, G.~E. 2010, in Proceedings of the 27th International
  Conference on Machine Learning (ICML-10), June 21-24, 2010, Haifa, Israel,
  807--814

\bibitem[{{Pesnell} {et~al.}(2012){Pesnell}, {Thompson}, \&
  {Chamberlin}}]{sdo2012}
{Pesnell}, W.~D., {Thompson}, B.~J., \& {Chamberlin}, P.~C. 2012, \solphys,
  275, 3

\bibitem[{Peyrard {et~al.}(2015)Peyrard, Mamalet, \& Garcia}]{Peyrard15}
Peyrard, C., Mamalet, F., \& Garcia, C. 2015, in VISAPP (1), ed. J.~Braz,
  S.~Battiato, \& F.~H. Imai (SciTePress), 84--91

\bibitem[{{Pietarila} {et~al.}(2013){Pietarila}, {Bertello}, {Harvey}, \&
  {Pevtsov}}]{Pietarila2013}
{Pietarila}, A., {Bertello}, L., {Harvey}, J.~W., \& {Pevtsov}, A.~A. 2013,
  \solphys, 282, 91

\bibitem[{{Quintero Noda} {et~al.}(2015){Quintero Noda}, {Asensio Ramos},
  {Orozco Su{\'a}rez}, \& {Ruiz Cobo}}]{Quintero2015}
{Quintero Noda}, C., {Asensio Ramos}, A., {Orozco Su{\'a}rez}, D., \& {Ruiz
  Cobo}, B. 2015, \aap, 579, A3

\bibitem[{{Richardson}(1972)}]{richardson72}
{Richardson}, W.~H. 1972, Journal of the Optical Society of America
  (1917-1983), 62, 55

\bibitem[{{Ruiz Cobo} \& {Asensio Ramos}(2013)}]{ruizcobo_asensioramos13}
{Ruiz Cobo}, B. \& {Asensio Ramos}, A. 2013, \aap, 549, L4

\bibitem[{Rumelhart {et~al.}(1988)Rumelhart, Hinton, \&
  Williams}]{Rumelhart1988}
Rumelhart, D.~E., Hinton, G.~E., \& Williams, R.~J. 1988 (Cambridge, MA, USA:
  MIT Press), 696--699

\bibitem[{{Schawinski} {et~al.}(2017){Schawinski}, {Zhang}, {Zhang}, {Fowler},
  \& {Santhanam}}]{schawinski17}
{Schawinski}, K., {Zhang}, C., {Zhang}, H., {Fowler}, L., \& {Santhanam}, G.~K.
  2017, \mnras, 467, L110

\bibitem[{Scherrer {et~al.}(2012)Scherrer, Schou, Bush, Kosovichev, Bogart,
  Hoeksema, Liu, Duvall, Zhao, Title, Schrijver, Tarbell, \&
  Tomczyk}]{Scherrer2012}
Scherrer, P.~H., Schou, J., Bush, R.~I., {et~al.} 2012, Solar Physics, 275, 207

\bibitem[{{Schmidhuber}(2014)}]{Overview2014}
{Schmidhuber}, J. 2014, ArXiv e-prints [\eprint[arXiv]{1404.7828}]

\bibitem[{{Shi} {et~al.}(2016){Shi}, {Caballero}, {Husz{\'a}r}, {Totz},
  {Aitken}, {Bishop}, {Rueckert}, \& {Wang}}]{Shi2016}
{Shi}, W., {Caballero}, J., {Husz{\'a}r}, F., {et~al.} 2016, ArXiv e-prints
  [\eprint[arXiv]{1609.05158}]

\bibitem[{{Simonyan} \& {Zisserman}(2014)}]{veryDeep2014}
{Simonyan}, K. \& {Zisserman}, A. 2014, ArXiv e-prints
  [\eprint[arXiv]{1409.1556}]

\bibitem[{{Socas-Navarro}(2005)}]{socas05}
{Socas-Navarro}, H. 2005, \apj, 621, 545

\bibitem[{{Stein}(2012)}]{stein12_a}
{Stein}, R.~F. 2012, Living Reviews in Solar Physics, 9, 4

\bibitem[{{Stein} \& {Nordlund}(2012)}]{stein12_b}
{Stein}, R.~F. \& {Nordlund}, {\AA}. 2012, \apjl, 753, L13

\bibitem[{{Tadesse} {et~al.}(2013){Tadesse}, {Wiegelmann}, {Inhester},
  {MacNeice}, {Pevtsov}, \& {Sun}}]{Tadesse2013}
{Tadesse}, T., {Wiegelmann}, T., {Inhester}, B., {et~al.} 2013, \aap, 550, A14

\bibitem[{Tai {et~al.}(2017)Tai, Yang, \& Liu}]{Ying2017}
Tai, Y., Yang, J., \& Liu, X. 2017, in In Proceeding of IEEE Computer Vision
  and Pattern Recognition, Honolulu, HI

\bibitem[{Tipping \& Bishop(2003)}]{Tipping03}
Tipping, M.~E. \& Bishop, C.~M. 2003, in Advances in Neural Information
  Processing Systems (MIT Press), 1303--1310

\bibitem[{{Tsuneta} {et~al.}(2008){Tsuneta}, {Ichimoto}, {Katsukawa}, {Nagata},
  {Otsubo}, {Shimizu}, {Suematsu}, {Nakagiri}, {Noguchi}, {Tarbell}, {Title},
  {Shine}, {Rosenberg}, {Hoffmann}, {Jurcevich}, {Kushner}, {Levay}, {Lites},
  {Elmore}, {Matsushita}, {Kawaguchi}, {Saito}, {Mikami}, {Hill}, \&
  {Owens}}]{tsuneta_hinode08}
{Tsuneta}, S., {Ichimoto}, K., {Katsukawa}, Y., {et~al.} 2008, \solphys, 249,
  167

\bibitem[{{van Noort}(2012)}]{vannoort12}
{van Noort}, M. 2012, A\&A, 548, A5

\bibitem[{{V{\"o}gler} {et~al.}(2005){V{\"o}gler}, {Shelyag}, {Sch{\"u}ssler},
  {Cattaneo}, {Emonet}, \& {Linde}}]{vogler05}
{V{\"o}gler}, A., {Shelyag}, S., {Sch{\"u}ssler}, M., {et~al.} 2005, \aap, 429,
  335

\bibitem[{{Wachter} {et~al.}(2012){Wachter}, {Schou}, {Rabello-Soares},
  {Miles}, {Duvall}, \& {Bush}}]{Wachter2012}
{Wachter}, R., {Schou}, J., {Rabello-Soares}, M.~C., {et~al.} 2012, \solphys,
  275, 261

\bibitem[{Xu {et~al.}(2014)Xu, Ren, Liu, \& Jia}]{Xu2014}
Xu, L., Ren, J. S.~J., Liu, C., \& Jia, J. 2014, in Proceedings of the 27th
  International Conference on Neural Information Processing Systems, NIPS'14
  (Cambridge, MA, USA: MIT Press), 1790--1798

\bibitem[{{Yeo} {et~al.}(2014){Yeo}, {Feller}, {Solanki}, {Couvidat},
  {Danilovic}, \& {Krivova}}]{Yeo2014}
{Yeo}, K.~L., {Feller}, A., {Solanki}, S.~K., {et~al.} 2014, \aap, 561, A22

\bibitem[{{Zhao} {et~al.}(2017){Zhao}, {Wang}, {Dong}, {Jia}, {Yang}, {Liu}, \&
  {Gao}}]{Zhao2017}
{Zhao}, Y., {Wang}, R., {Dong}, W., {et~al.} 2017, ArXiv e-prints
  [\eprint[arXiv]{1703.04244}]

\end{thebibliography}

\end{document}